\documentclass[12pt]{article}
\usepackage[top=1.0in, bottom=1.0in, left=1.12in, right=1.12in]{geometry}

\usepackage{graphicx} 
\usepackage{caption}
\usepackage{subcaption}
\usepackage{algorithm}
\usepackage{algorithmic}
\usepackage{hyperref}
\usepackage{amsmath, amssymb, amsthm}
\usepackage{natbib}

\usepackage{color}
\usepackage{booktabs}

\usepackage{setspace}
\newcommand{\beq}{\vspace{0mm}\begin{equation}}
\newcommand{\eeq}{\vspace{0mm}\end{equation}}
\newcommand{\beqs}{\vspace{0mm}\begin{eqnarray}}
\newcommand{\eeqs}{\vspace{0mm}\end{eqnarray}}
\newcommand{\barr}{\begin{array}}
\newcommand{\earr}{\end{array}}

\newcommand{\Nmat}[0]{{{\bf N}}}

\newcommand{\nv}[0]{{\boldsymbol{n}}}

\newcommand{\cdotv}{\boldsymbol{\cdot}}

\newcommand{\given}{\,|\,}

\newcommand{\thetav}{\boldsymbol{\theta}}

\newcommand{\piv}{\boldsymbol{\pi}}

\newcommand{\E}{\mathbb{E}}

\usepackage{natbib} 

\renewcommand\footnotemark{}

\begin{document}
\title{BNP-Seq: Bayesian Nonparametric Differential Expression Analysis of Sequencing Count Data}
 \author{ 
 Siamak Zamani Dadaneh$^*$,\,\, Xiaoning Qian$^*$$^{\#}$,\,\, and Mingyuan Zhou$^\dagger$$^{\#}$\\
 \thanks{\newline 
 $^*$ Department of Electrical \& Computer Engineering, Center for Bioinformatics \& Genomic Systems Engineering,
 Texas A\&M University, College Station, TX 77843, USA.\newline
 $^\dagger$ Department of Information, Risk, \& Operations Management and Department of Statistics and Data Sciences, The University of Texas at Austin, Austin, TX 78712, USA. \newline
 $^{\#}$ Correspondence should be addressed to Xiaoning Qian (\texttt{xqian@ece.tamu.edu}) or Mingyuan Zhou (\texttt{mingyuan.zhou@mccombs.utexas.edu}).}
}

\begin{spacing}{1.25}
\maketitle
\begin{abstract}

We perform differential expression analysis of high-throughput sequencing count data under a Bayesian nonparametric framework, removing sophisticated \emph{ad-hoc} pre-processing steps commonly required in existing algorithms. We propose to use the gamma (beta) negative binomial process, which takes into account different sequencing depths using sample-specific negative binomial probability (dispersion) parameters, to detect differentially expressed genes by comparing the posterior distributions of gene-specific negative binomial dispersion (probability) parameters. These model parameters are inferred by borrowing statistical strength across both the genes and samples. Extensive experiments on both simulated and real-world RNA sequencing count data show that the proposed differential expression analysis algorithms clearly outperform previously proposed ones in terms of the areas under both the receiver operating characteristic and precision-recall curves.

\vspace{4mm}
\noindent%

\end{abstract}

\end{spacing}

\section{Introduction}
\label{sec:intro}

There has been significant recent interest in analyzing RNA sequencing~(RNA-Seq) count data for studying life systems~\citep{wang2009rna, metzker2010sequencing}. 
It is challenging to model RNA-Seq data, not only because it is typically a large-$p$-small-$n$ problem \citep{West03bayesianfactor} where the data dimension is high while the sample size is small, but also because the sequencing counts are 
nonnegative, skewed, having large dynamical ranges, and highly over-dispersed 
 \citep{deseq,datta2014statistical}. A key task in RNA-Seq analysis 
is to identify the genes that are differentially expressed 
between different groups of samples ($e.g.$, samples measured under different medical conditions) \citep{wang2010degseq,robinson2010scaling, 
deseq,
oshlack2010rna,li2011normalization,li2013finding}. The expression level of each RNA locus, here the gene, is determined by the number of sequenced reads to the transcript \citep{mortazavi2008mapping}. Unlike a gene probe based method such as microarrays \citep{schena1995quantitative}, the abundance of genes in RNA-Seq is restricted by the sequencing depth and there often exist dependencies between the expressions of different transcripts \citep{roberts2011improving}.

Modeling the sequencing counts using an over-dispersed count distribution, such as the negative binomial (NB) distribution \citep{Yule,
NB_Fitting_53}, is one of the most popular approaches for differential expression analysis \citep{edger,deseq}. 
In the null hypothesis that a gene is not differently expressed, it is common to assume that the expectations of the counts of that gene are the same across different groups, after making adjustments to account for both technical and biological variations. In particular, almost all existing 
comparative analysis algorithms, before 
downstream analyses, require 
normalizing the sequencing counts 
to compensate the variations of 
sequencing depths across samples \citep{soneson2013comparison,dillies2013comprehensive,zyprych}. For instance, edgeR and DESeq, 
two widely used differential expression analysis R software packages adopt different \emph{ad-hoc} normalization procedures: 
edgeR either calculates a trimmed mean of M-values \citep{robinson2010scaling} between each pair of samples or uses an upper quantile of samples \citep{bullard2010evaluation} for normalization \citep{edger}, while DESeq takes the median of the ratios of observed sample's counts to the geometric mean across samples as a scaling factor for that specific sample \citep{deseq,deseq2}.

Normalizing the sequencing counts, however, inevitably destroys the discrete nature of the raw data and makes the performance clearly depend on whether the introduced normalization is suitable for 
the structure of the RNA-Seq data under study \citep{soneson2013comparison,dillies2013comprehensive,zyprych}. If the normalization procedure extracts normalization constants from the data under study to parameterize the distributions of the gene counts, the discrete nature of the raw data is preserved, but the model can no longer be considered as a generative model. In addition, almost all existing normalization procedures assume that most of the genes are not differentially expressed, and the differentially expressed genes are equally likely to be up- and down-regulated \citep{loven2012revisiting,lorenz2014using,risso2014normalization,risso2014role}. The violation of the assumption may potentially be addressed by using external RNA control consortium (ERCC) spike-in sequences for controls; however, it is shown in \citet{risso2014normalization,risso2014role} that the read counts for ERCC spike-ins alone are usually not stable enough to be used for normalization. 
Moreover, despite that a wide array of methods 
have been proposed to adjust the counts to account for technical and biological variations, 
there is not a single one that clearly outperforms the others 
under various scenarios 
\citep{soneson2013comparison,comprehensive,dillies2013comprehensive,risso2014normalization,zyprych,zhang2014comparative}.

In this paper, rather than heuristically modifying the data 
and estimating the model 
parameters based on an empirical Bayes procedure, 
we introduce a 
generative model to analyze differential expression directly on the raw sequencing counts, without the need to preprocess the data by normalization. Instead of using parametric count distributions to describe the counts, we use a stochastic process to model the observed sample-gene random count matrix in each group, whose model parameters are estimated by sharing statistical strength across both the genes and samples. The stochastic process can be used to explain not only the counts and the total number of expressed genes in the observed random count matrix, but also the number of newly expressed genes and the counts on both existing and newly expressed genes to be brought by a new sample. Such flexible random-process-based models lift the need of \emph{ad-hoc} data normalization and strict parametric assumptions, allowing heterogeneity across samples and gene expression variations across different conditions to be well captured. 

More specifically, moving beyond existing algorithms that model over-dispersed counts with the NB distribution, our Bayesian nonparametric (BNP) algorithms model the gene counts using the gamma-negative binomial 
process (GNBP) \citep{NBP_CountMatrix}, which mixes the NB shape parameter for each gene with the distribution of the weight of an atom of a gamma process \citep{ferguson73}, 
or beta-negative binomial 
 process (BNBP) \citep{BNBP_PFA_AISTATS2012,NBP2012,NBPJordan}, which mixes the NB probability parameter of each gene with the distribution of the weight of an atom of a beta process \citep{Hjort}. 
In addition to the GNBP and BNBP, for comparison, we have 
extended the negative binomial process (NBP) of \citet{NBP_CountMatrix} by multiplying the gene-specific Poisson rates with gamma distributed sample-specific scaling parameters, and refer to it as the scaled NBP. 
While the NBP of \citet{NBP_CountMatrix} is not expected to work well since it does not explicitly model the variation of a sample's total count, the scaled NBP, even with a scaling parameter for each sample to capture that variation, is found to provide poor performance, indicating a clear limitation of the Poisson distribution assumption. 
We will show that while the variations of 
the gene counts across samples are well captured by neither the Poisson rates of the scaled NBP nor the normalized Poisson rates of the NBP, 
they are well modeled by both the GNBP and BNBP, using the NB shape and probability parameters, respectively.

Unlike previous algorithms for differential expression analysis, the proposed BNP algorithms require no normalization pre-processing steps and they infer the posterior distributions, instead of point estimates, of their model parameters, using 
Gibbs sampling with closed-form update equations, achieving state-of-the-art performance in detecting truly differentially expressed genes for both synthetic and real data. 
To 
our knowledge, we are the first in 
constructing BNP algorithms with these distinct advantages to analyze the sequencing counts to detect differential expression in genomics.

The remainder of the paper is organized as follows. After reviewing existing differential expression analysis algorithms in Section \ref{sec:1.1}, we introduce the NBP-Seq, GNBP-Seq, and BNBP-Seq for differential expression analysis in Section~\ref{sec:methods}.  
We present experimental results on both synthetic and real-world benchmark RNA-Seq data in Section~\ref{sec:3} and show that both the proposed GNBP-Seq and BNBP-Seq clearly outperform previously proposed differential express analysis algorithms. We conclude the paper in Section~\ref{sec:conc}. 

\section{Differential expression analysis} \label{sec:1.1}
For $J$ RNA-Seq samples organized into the same group, let us denote $n_{jk}$ as the number of reads in sequencing sample $j\in\{1,\ldots,J\}$ that are assigned to gene $k\in\{1,\ldots,K\}$, where $K$ is the number of genes in the genome. 
Since the counts of a gene across samples are often over-dispersed, it is natural to model them using a NB distribution, where its variance $\sigma^2$ is related to its mean $\mu$ as $\sigma^2=\mu+\phi \mu^2$, where $\phi$ is the dispersion parameter. As it is also common to refer to $r=\phi^{-1}$ as the dispersion parameter, to avoid ambiguity, we will refer to $r=\phi^{-1}$ as the NB shape parameter. 

Methods such as edgeR and DESeq propose different ways to estimate $\phi$. EdgeR models the gene count $n_{jk}$ as a NB distribution with mean $n_j \lambda_{jk}$ and dispersion $\phi_{k}$, where $n_j$ is the observed total count (or the sum of adjusted counts) for sample $j$, $\lambda_{jk}$ represents the abundance of gene $k$ in sample $j$, and $\phi_k$ is considered as the coefficient of biological variation that is estimated by conditional maximum likelihood \citep{smyth1996conditional}. Furthermore, an empirical Bayes procedure is applied to shrink the dispersion parameters $\phi_k$ towards a common value \citep{robinson2007}.

DESeq also models 
the gene counts with the NB distribution. It considers two terms to estimate the variance $\sigma_{jk}^2$ for gene $k$ in sample $j$, where the first term (shot noise) is associated with the mean expression of the gene, and the second one (raw variance) takes into account the biological variations between replicates. More specifically, it lets $\sigma_{jk}^2 = \mu_{jk} + n_j^2 v_{k,\rho(j)}$. Here, $\rho(j)$ is the group to which sample $j$ belongs, and $v_{k,\rho(j)}$ is the per-gene raw variance, which is a smooth function of $\lambda$ and $\rho$, an assumption that allows pooling data from different genes to estimate their variances.

Another widely used tool, baySeq \citep{bayseq}, takes an empirical Bayesian approach to estimate the posterior probabilities 
of a set of models that define different patterns of differential expression for each gene. For instance, in the simplest case of a pairwise comparison between conditions A and B, with two biological replicates for each condition, the model for no differential expression is defined by the set of samples \{A1, A2, B1, B2\}, while differential expression between conditions A and B is defined by the sets \{A1, A2\} and \{B1, B2\}. The method then assumes that the counts follow the NB distribution and derives an empirically determined prior distribution from the data.

The final component of these methods is the test for gene differential expression. Both edgeR and DESeq use variations of Fisher's exact test, adjusted for the NB distribution, to compute exact $p$-values for the null hypothesis that the mean expressions of the genes are equal in both conditions under comparison. EdgeR also considers the generalized linear model approach to identify differentially expressed genes in its later versions; nevertheless, it has been shown to have similar performance to the method based on Fisher's exact test \citep{schurch2016many}. Different from edgeR and DESeq, baySeq ranks the genes based on the inferred posterior probabilities of differential expression.

\section{Bayesian nonparametric differential expression analysis for RNA-Seq} \label{sec:methods}

We consider a family of NB processes, each of which can be used to 
describe the row-by-row sequential construction of a sample-gene sequencing count matrix, where the addition of a new sample (row) brings counts at not only previously expressed genes (columns), but also previously unexpressed ones. We also describe the equivalent construction that draws a Poisson random number of independent, and identically distributed~(i.i.d.) columns simultaneously, where each column corresponds to the counts of a gene that is expressed at least once across all the observed samples of a group. Showing these two equivalent constructions helps clearly understand the underlying statistical assumption made on the RNA-Seq data by a BNP prior, and how the statistical strength is shared across both the genes and samples to estimate both the sample-specific model parameters, which account for the variations in sequencing depths, and the gene-specific model parameters, whose posterior distributions are used to detect differentially expressed genes.

We explore a family of NB processes, specifically NBP, GNBP, and BNBP in this paper. 
To model the gene counts in each group with the GNBP, we consider the null hypothesis that with sample-specific NB probability parameters, the posterior distributions of the gene-specific NB shape parameters, regularized by the gamma process in the prior, are the same across different groups; whereas with the BNBP, we consider the null hypothesis that with sample-specific NB shape parameters, the posterior distributions of the gene-specific NB probability parameters, regularized by the beta process in the prior, are the same across different groups. For comparison, we also include NBP and its scaled version, and consider the null hypothesis that the posterior distributions of the gene-specific Poisson rate parameters of the scaled NBP, regularized by the gamma process in the prior, or the corresponding normalized Poisson rates of the NBP, are the same across different groups. Instead of following a standard hypothesis testing procedure to test whether two parameters are equal, 
in this paper, after fitting the sequencing counts with a BNP prior, we assess the significance of gene expression changes across groups by measuring the distances between the gene-specific posterior distributions of the NB shape parameters for the GNBP, NB probability parameters for the BNBP, Poisson rates for the scaled NBP, and normalized Poisson rates for the NBP, using symmetric Kullback-Leibler (KL) divergence \citep{kullback1951information}. Note that there are two common ways to parameterize a NB distribution: either via both its mean $\mu$ and shape parameter $r$, or via both its shape parameter $r$ and probability parameter $p$. These two different ways are related in that $\mu=rp/(1-p)$ and the variance can be expressed as $\mu+\mu^2/r = rp/(1-p)^2=\mu/(1-p)$. We  parameterize a NB distribution with both its shape and probably parameters throughout the paper. 

Below we show how a stochastic process can be used to model the counts in each group, where the group index is omitted for brevity. We represent the counts of all expressed genes in a group as a random count matrix 
$\Nmat_J\in \mathbb{Z}^{J \times K_J}$, where $\mathbb{Z}=\{0,1,\ldots\}$ represents the set of nonnegative integers, $K_J$ denotes the random number of genes that are expressed at least once in the $J$ samples of the group, and 
the element $n_{jk}$ 
represents the number of reads in sequencing sample $j\in\{1,\ldots,J\}$ that are assigned to gene $k\in\{1,\ldots,K_J\}$. Note that $K_J$, the number of expressed genes among the $J$ samples, is smaller or equal to $K$, the total number of genes in the genome, and $K_J$ can potentially increase without bound as $J$ increases. For graphical illustrations of random count matrices generated by the NBP, GNBP, and BNBP, we refer the interested readers to Figures 1 and 2 of \citet{NBP_CountMatrix}.

\subsection{NBP-Seq: Negative binomial process for RNA-Seq}

Let us denote $G_0$ as a finite and continuous base measure over a complete and separable metric space $\Omega$, $c\in\mathbb{R}_+$ as a scale parameter, and $q_j\in\mathbb{R}_+$ as sample-specific scaling parameters, where $\mathbb{R}_+:=\{x:x>0\}$. 
We define the scaled 
negative binomial process~(NBP) that has sample-specific scaling parameters as
$$
(X_1,\ldots,X_J) \given c, G_0, \{q_j\}_{1,J} \sim\mbox{NBP}(G_0,c_0,q_1,\ldots,q_J),
$$ 
which is obtained by marginalizing out a gamma process \citep{ferguson73} $G \sim \Gamma\mbox{P}(G_0, 1/c)$ from $J$ conditionally independent Poisson processes \citep{PoissonP} $X_j\given {q}_j, G \sim \mbox{PP}({q}_j G)$, where for disjoint Borel sets $A_j\subset \Omega$, the gamma process $G$ is defined such that $G(A_i)\sim\mbox{Gamma}[G_0(A_i), 1/c]$ are independent gamma random variables, and the Poisson process $X_j$ is defined such that $X_j(A_i) \sim \mbox{Pois}[{q}_j G(A_i)]$ are independent Poisson random variables. With a draw from the gamma process expressed as $G=\sum_{k=1}^{\infty} r_k \delta_{\omega_k}$, where $\omega_k$ and $r_k$ are the atoms and their weights, respectively, a draw from $X_j$ can be expressed as 
\beq
X_j=\sum_{k=1}^{\infty} n_{jk} \delta_{\omega_k},~ n_{jk} \sim \mbox{Pois}({q}_j r_k).
\eeq 
Note that if we fix $q_j=1$ for all $j$, then the proposed NBP with sample-specific scaling parameters reduces to the NBP in \citet{NBP2012} and \citet{NBP_CountMatrix}.

The conditional likelihood of the observed $J$ samples of a group can be written as
\beq\label{eq:clnbp}
p(\{X_j\}_{j=1}^J\given G)= e^{-{q}_{\cdotv}G(\Omega \setminus \mathcal{D}_J)} \left[\prod_{k=1}^{K_J} \frac{r_k^{n_{\cdotv k}} e^{-{q}_{\cdotv}r_k}}{ \prod_{j=1}^{J} n_{jk}! }\right]\left[\prod_{j=1}^J q_j^{n_{j}}\right],
\eeq
where $\mathcal{D}_J=\{\omega_k\}_{k:n_{\cdotv k}>0}$ is the set of points of discontinuity, $K_J=|\mathcal{D}_J| = \sum_{k} \delta(n_{\cdotv k}>0)$ is the number of genes that are expressed at least once, ${q}_{\cdotv} = \sum_{j=1}^J {q}_j$, and $n_{\cdotv k} = \sum_{j=1}^{J} n_{jk}$. We map the counts associated with the elements of $\mathcal{D}_J$ 
to the random count matrix $\Nmat_J$. While the labelings of the atoms in $\mathcal{D}_J$ are arbitrary, they are mapped in one of the $K_J!$ possible ways to the columns of $\Nmat_J$. Similar to the derivation in \citet{NBP_CountMatrix},
using a marginalization procedure shown in \citet{CarTehMur2013a}, 
one may marginalize out the gamma process $G$, leading to the distribution of the random count matrix as
\begin{align}\label{eq:NBP_PMF}
f(\Nmat_J \mid \gamma_0,c,q_{1},\ldots,q_J) &=\frac{p(\{X_j\}_{1,J} \mid \gamma_0, c,q_{1},\ldots,q_J)}{K_J!} \notag\\
&= \frac{{\gamma_0^{K_J} \exp\left[-\gamma_0\ln(\frac{{q}_{\cdotv}+c}{c}) \right] }}{K_J!}\left[ \prod_{k=1}^{K_J} \frac{ \frac{\Gamma(n_{\cdotv k})}{({q}_{\cdotv}+c)^{n_{\cdotv k}}}}{\prod_{j=1}^J n_{jk}!}\right] \left[\prod_{j=1}^J q_j^{n_{j}}\right] \, .
\end{align}

One may verify by straightforward calculation that a scaled NBP random count matrix with the probability mass function~(PMF) shown in (\ref{eq:NBP_PMF}) can be generated column by column as i.i.d. count vectors:
\begin{align}\label{eq:NBProwwise}
&\nv_{:k} \sim \mbox{Multinomial}(n_{\cdotv k}, {q}_1/{q}_{\cdotv},\ldots, {q}_J/{q}_{\cdotv}),\notag \\ 
&n_{\cdotv k}\sim\mbox{Logarithmic}[{ {q}_{\cdotv}}/{(c+ {q}_{\cdotv})}],\notag\\
& K_J\sim \mbox{Pois}\left\{\gamma_0\left[\ln(c+{q}_{\cdotv})-\ln(c)\right]\right\}\, .
\end{align}
It is clear from (\ref{eq:NBProwwise}) that the columns of $\Nmat_J$ are 
i.i.d. multivariate count vectors, which all follow the same logarithmic-multinomial (mixture) distribution. Thus the scaled NBP random count matrix $\Nmat_J$ is column exchangeable. It is also row exchangeable if and only if the $q_j$ are the same for all $j\in\{1,\ldots,J\}$.

Now consider the row-wise sequential construction of the scaled NBP random matrix. 
With the prior on $\Nmat_J\in\mathbb{Z}^{J\times K_J}$ well defined, 
straightforward calculations using (\ref{eq:NBProwwise}) 
yield the following form for this prediction rule, 
expressed in terms of familiar PMFs: 
\begin{align}
\label{eq:NBP_pre}\small
 \frac{f(\Nmat_{J+1}\!\mid \!\thetav)}{f(\Nmat_{J}\!\mid\! \thetav)} &= \frac{K_J! K^+_{J+1}!}{K_{J+1}!}  \prod_{k=1}^{K_J}\mbox{NB}
\left(n_{(J+1)k};n_{\cdotv k},\frac{{q}_{J+1}}{c+{q}_{\cdotv}+{q}_{J+1}}\right)\notag\\
&\times \prod_{k=K_J+1}^{K_{J+1}} \mbox{Logarithmic}\left(n_{(J+1)k};\frac{{q}_{J+1}}{c+ {q}_{\cdotv}+{q}_{J+1}}\right)\notag\\
&\times \mbox{Pois}\left\{K^+_{J+1}; \gamma_0\left[\ln(c+{q}_{\cdotv}+{q}_{J+1})-\ln(c+{q}_{\cdotv})\right]\right\},
\end{align}
where $\thetav:=\{\gamma_0,c,q_1,\ldots,q_J\}$.
This formula indicates that, to add a new row 
to $\Nmat_J\in\mathbb{Z}^{J \times K_J}$, 
we first draw count $\mbox{NB}[n_{\cdotv k},{q}_{J+1}/(c+{q}_{\cdotv}+{q}_{J+1})]$ at each existing column. We then draw $K^+_{J+1}$ new columns as
$K^+_{J+1}\sim\mbox{Pois}\{ \gamma_0\left[\ln(c+{q}_{\cdotv}+{q}_{J+1})-\ln(c+{q}_{\cdotv})\right]\}$. Finally, each entry in the new columns has a $\mbox{Logarithmic}\left[n_{(J+1)k};{{q}_{J+1}}/({c+ {q}_{\cdotv}+{q}_{J+1}})\right]$ distributed random count. 
It is clear in the sequential construction of the scaled NBP random count matrix, for a point of discontinuity $\omega_k\in\mathcal{D}_J$, the variance and mean are related as
\beq
 \mbox{var}[n_{(J+1)k}] = \E[n_{(J+1)k}]+\frac{\E^2[n_{(J+1)k}]}{n_{\cdotv k}}. \label{eqn:varmeanNBP} 
\eeq
Since $n_{\cdotv k}$, the total count of gene $k$ of all the $J$ samples of the group, is fixed, the above equation indicates 
a variance and mean relationship that does not change. 

\subsubsection{Inference for the scaled NBP}

The parameters of the scaled NBP can be inferred using Gibbs sampling with closed-form update equations. 
Using likelihoods (\ref{eq:clnbp}) and (\ref{eq:NBP_PMF}), with $\gamma_0 \sim \text{Gamma}(e_0,1/f_0)$, $c \sim \text{Gamma}(c_0,1/d_0)$, and $q_j\sim \text{Gamma}(a_0,1/b_0)$ in the prior, each Gibbs sampling iteration proceeds as 
\begin{align}\label{eq:gibbsNBP}
&(\gamma_0\given-) \sim \text{Gamma} \Big( e_0+K_J, \frac{1}{f_0 - \ln(\frac{c}{c+{q}_{\cdotv}})} \Big),\nonumber\\
&(r_k\given-) \sim \text{Gamma} [n_{\cdotv k}, 1/(c+{q}_{\cdotv})], \nonumber\\
&[{G(\Omega \setminus \mathcal{D}_J)\given-} ]\sim \text{Gamma} [\gamma_0, 1/(c+{q}_{\cdotv})], \nonumber\\
&(q_j\given-)\sim \text{Gamma}\{a_0+n_j,1/[b_0+G(\Omega)]\}, \nonumber\\
&(c\given-) \sim \text{Gamma} \{c_0 + \gamma_0, 1/[d_0+G(\Omega)]\},
\end{align}
where $G(\Omega):=G(\Omega\backslash\mathcal{D}_J) + \sum_{k=1}^{K_J} r_k$, given which the total gene count for sample $j$ follows $\mbox{Poisson}[q_jG(\Omega)]$. Note that a gene that has at least one nonzero count among the $J$ samples will be attached to a discrete atom (point of discontinuity) of the gamma process with weight $r_k$, while all the other countably infinite unexpressed genes are associated with the atoms in the absolute continuous space $\Omega\backslash \mathcal{D}_J$, whose total weight is $G(\Omega\backslash \mathcal{D}_J)$.  

\subsubsection{NBP-Seq differential expression analysis}

To detect differentially expressed genes using the scaled NBP, we notice in the prior that
$$\E[n_{jk}\given q_j, G]=\mbox{var}[n_{jk}\given q_j,G] = q_jr_k$$
and in the conditional posterior shown in \eqref{eq:gibbsNBP} that
\beq\label{eq:E_NBP}
\E[r_k\given-] = n_{\cdotv k}/(c+{q}_{\cdotv}),~~~~~\E[q_j\given-] = (a_0+n_j)/[b_0+G(\Omega)].
\eeq
Thus one may consider $r_k$ as a gene-specific Poisson rate parameter that indicates the expression level of gene $k$, whose conditional posterior 
is related to both $n_{\cdotv k}$, the total count of gene $k$ across all the $J$ samples of the group, and $q_{\cdotv}$, the total sum of the sample-specific gamma distributed scaling parameters; one may consider $q_j$ as a scaling factor to be inferred from the data, whose conditional posterior is determined not only by $n_j$, the total count of all genes in sample $j$ that indicates the sequencing depth of sample~$j$, but also by $G(\Omega)$, 
the total sum of all countably infinite gene-specific Poisson rate parameters; and the conditional posterior of $\gamma_0$ is clearly related to $K_J$, the total number of expressed genes in the group. Therefore, the scaled NBP borrows statistical strength across both the genes and samples to infer the conditional posterior of $r_k$. %

To assess whether the difference between the expressions of the same gene at different sample groups is statistically significant, we collect posterior Markov chain Monte Carlo~(MCMC) samples for each $r_k$ in each group, and use these MCMC samples to measure the distance between the posterior distributions of the $r_k$ of the same gene across different groups. Note that for a gene whose total count across all samples in a group is zero, the posterior values of its $r_k$ would be fixed at 0. 

Instead of using the scaled NBP that introduces $q_j$ to model sample-specific sequencing depths, we also consider the original NBP of \citet{NBP_CountMatrix} with all $q_j$ fixed at one. To compensate for the variations of sequencing depths between samples, for the original NBP, we normalize the inferred Poisson rates $r_k$ and use them to evaluate the significance of differential gene expressions.

\subsection{GNBP-Seq: Gamma-negative binomial process for $\text{RNA-Seq}$}

To generate the 
random count matrix $\Nmat_J$ in a group, 
we construct a gamma-negative binomial process (GNBP) \citep{NBP_CountMatrix} as 
\begin{equation}
X_j\given G \sim \mbox{NBP}(G,p_j), \;\;\; G \sim \Gamma \mbox{P}(G_0,1/c),
\end{equation}
where $j\in\{1,..,J\}$ and $X_j\given G \sim \mbox{NBP}(G,p_j)$ is defined as a NBP such that $X_j(A) \sim \mbox{NB}[G(A),p_j]$ for each Borel subset $A \subset \Omega$. 
Note that $X_j\given G \sim \mbox{NBP}(G,p_j)$ can also be augmented as a gamma process mixed sum-logarithmic process~(SumLogP) as
\begin{equation}
X_j\given L_j \sim \mbox{SumLogP}(L_j,p_j),~ L_j \given G\sim \mbox{PP}(q_j G), 
\end{equation}
where $q_j:=-\ln(1-p_j)$, $i.e.$, $p_j=1-e^{-q_j}$, and the SumLogP is defined in \citet{NBP_CountMatrix} such that $X_j(A) = \sum_{t=1}^{L_j(A)} u_t, ~u_t\sim\mbox{Logarithmic}(p_j)$ for each Borel subset $A \subset \Omega$. Thus the GNBP also can be expressed as a NBP mixed SumLogP as
\begin{equation}
X_j\given L_j \sim \mbox{SumLogP}(L_j,p_j),~(L_1,\ldots,L_J) 
 \sim\mbox{NBP}(G_0,c,q_1,\ldots,q_J).
\end{equation}

With a draw from the gamma process $G$ expressed as $G=\sum_{k=1}^{\infty} r_k \delta_{\omega_k}$, a draw from $X_j$ can be expressed as 
\beq
X_j=\sum_{k=1}^{\infty} n_{jk} \delta_{\omega_k},~ n_{jk} \sim \mbox{NB}(r_k,p_j).
\eeq
The GNBP employs sample-specific NB probability parameters $p_j$ to model row heterogeneity. In the context of RNA-Seq data, 
the variations of $p_j$ 
can be used to account for those of sequencing depths. 

Both the row-wise and column-wise constructions of the GNBP random count matrix mimic these of the NBP random count matrix. They are described in detail in \citet{NBP_CountMatrix} and hence omitted here for brevity. We mention that the two key differences in their row-wise sequential constructions are that the GNBP uses the gamma-NB instead of NB distributions to model the counts at previously expressed genes brought by a new sample, and the GNBP uses the logarithmic mixed sum-logarithmic instead of logarithmic distributions to model the counts at newly expressed genes brought by a new sample. 

As shown in \citet{NBP_CountMatrix}, 
in the sequential construction of the GNBP random count matrix, for a point of discontinuity $\omega_k\in\mathcal{D}_J$, the variance and mean are related as 
\beq
\mbox{var}[n_{(J+1)k}] = \frac{\E[n_{(J+1)k}]}{1-p_{J+1}} + \frac{ \E^2[n_{(J+1)k}]}{l_{\cdotv k}}, \label{eqn:varmeanGNBP} 
 \eeq
which depends on both $p_{J+1}$ and $l_{\cdotv k}$ that are random, where $l_{\cdotv k}:=\sum_{j=1}^J l_{jk},~l_{jk}\sim\mbox{CRT}(n_{jk},r_k)$, with the Chinese Restaurant Table~(CRT) distribution defined in the Appendix. 
Comparing \eqref{eqn:varmeanNBP} and \eqref{eqn:varmeanGNBP}, it is clear that since $p_{J+1}<1$ and $l_{\cdotv k}\le n_{\cdotv k}$, the GNBP can model much more over-dispersed counts than the NBP.

\subsubsection{Inference for the GNBP}
Letting $\gamma_0 \sim \text{Gamma}(e_0,1/f_0)$, $p_j \sim \text{Beta}(a_0,b_0)$, and $c \sim \text{Gamma}(c_0,1/d_0)$ in the prior, as in \citet{NBP_CountMatrix}, a Gibbs sampling iteration for the GNBP proceeds as
\begin{align}\label{eq:GNBP_sampling}
&(\gamma_0\given-)\sim\mbox{Gamma}\bigg(e_0+K_J, \frac{1}{f_0-\ln(\frac{c}{c+q_{\cdotv}})}\bigg),\notag\\
&(l_{jk}\given-)\sim\mbox{CRT}(n_{jk},r_k),
~(r_k\given-)\sim\mbox{Gamma}\big[ l_{\cdotv k}, 1/(c+q_{\cdotv})\big],\notag\\ &\{G(\Omega\backslash\mathcal{D}_J)\given-\}\sim\mbox{Gamma}\big[\gamma_0,1/(c+q_{\cdotv})\big],\notag\\ %
&(p_j\given-)\sim\mbox{Beta}\big[a_0+n_j,b_0 + G(\Omega)\big],\notag\\ &(c\given-)\sim\mbox{Gamma}\big\{c_0+\gamma_0,{1}/{[d_0 + G(\Omega)]\big\}}. \end{align}
Note that given $G(\Omega)$, the total gene count for sample $j$ follows $\mbox{NB}[G(\Omega),p_j]$.

\subsubsection{GNBP-Seq differential expression analysis}

In the GNBP, since in the prior we have $$\E[n_{jk}\given G,p_j] = r_k \frac{p_j}{1-p_j},$$ $$\mbox{var}[n_{jk}\given G,p_j] = r_k \frac{p_j}{(1-p_j)^2} = \E[n_{jk}\given G,p_j] + r_k^{-1}\E^2[n_{jk}\given G,p_j],$$ 
and in the conditional posterior, if $b_0 + G(\Omega)>1$, we have
\beq \label{eq:E_GNBP}
\E[r_k\given-] = l_{\cdotv k}/(c+q_{\cdotv}),~~~
\E[p_j/(1-p_j)\given-] = (a_0+n_j)/[b_0 + G(\Omega)-1].
\eeq
Thus one may interpret $p_j/(1-p_j)$ as a term that accounts for the sequencing depth of sample $j$, and may compare the posterior distributions of the NB shape parameter $r_k$ of the same gene at different groups to assess differential expression of that gene. The conditional posterior of the scaling factor $p_j/(1-p_j)$ is determined by not only $n_j$, the total counts of genes in sample $j$, but also $G(\Omega)$, the total sum of all countably infinite gene-specific NB shape parameters; and the conditional expectation of $r_k$ is related to both $l_{\cdotv}$ and $q_{\cdotv}$, which aggregate their corresponding sample-specific values across all the $J$ samples. Therefore, the GNBP borrows statistical strength across both the genes and samples to infer the conditional posterior of $r_k$. 
For an unexpressed gene, whose total count across all samples in a group is 0, the posterior values of its $r_k$ would be fixed at 0. 

Comparing \eqref{eq:E_NBP} and \eqref{eq:E_GNBP} shows that both the GNBP and scaled NBP have similar sample-specific scaling parameters, but, as in \eqref{eq:GNBP_sampling}, since $\E[l_{jk} \given -]= \sum_{t=1}^{n_{jk}} {r_k}/{(r_k+t-1)}$ and hence $\E[l_{jk} \given -] \approx r_k\ln(n_{jk}+r_k)$ for large $n_{jk}$, the posteriors of the gene-specific parameters $r_k$ in the GNBP would be 
impacted much less by some genes whose expressions $n_{jk}$ are significantly larger than their mean expression levels, which are commonly observed in genomic studies.

\subsection{BNBP-Seq: Beta-negative binomial process for RNA-Seq}

Similar to the GNBP, the BNBP can be used to model RNA-Seq samples. The BNBP can be constructed by sharing the NB probability parameters across the $J$ sequencing samples of the same group as
\begin{equation}
X_j\given r_j, B \sim \mbox{NBP}(r_j,B), \;\;\; B \sim \mbox{BP}(c,B_0),
\end{equation}
where $j\in\{1,\ldots,J\}$ and $B \sim \mbox{BP}(c,B_0)$ is a beta process with a finite and continuous base measure $B_0$ over $\Omega$ and a concentration parameter $c$, with L\'evy measure 
\begin{equation}
\nu (dpd\omega) = p^{-1} (1-p)^{c-1} dpB_0(d\omega).
\end{equation}
With a draw from the beta process $B$ expressed as $B=\sum_{k=1}^{\infty} p_k \delta_{\omega_k}$, where $\omega_k$ and $p_k$ are atoms and their associated probability weights, respectively, a draw from $X_j$ given $B$ can be expressed as 
\beq
X_j=\sum_{k=1}^{\infty} n_{jk} \delta_{\omega_k},~ n_{jk} \sim \mbox{NB}(r_j,p_k).
\eeq
In the BNBP, different $r_j$'s are used to model the sequencing depth variations.

Both the row-wise and column-wise constructions of the BNBP random count matrix, as described in detail in \citet{NBP_CountMatrix} and hence omitted here for brevity, mimic these of the scaled NBP random count matrix. We mention that the two key differences in their row-wise sequential constructions are that the BNBP uses the beta-NB instead of NB distributions to model the counts at previously expressed genes brought by a new sample, and the BNBP uses the digamma instead of logarithmic distributions to model the counts at newly expressed genes brought by a new sample. 

As shown in \citet{NBP_CountMatrix}, in the sequential construction of the BNBP random count matrix, for a point of discontinuity $\omega_k$, 
the variance and mean are related as 
\beq
\mbox{var}[n_{(J+1)k}] = \frac{\E[n_{(J+1)k}]}{\frac{c+r_{\cdotv} -2}{n_{\cdotv k}+c+r_{\cdotv} -1}} + \frac{ \E^2[n_{(J+1)k}]}{\frac{n_{\cdotv k}(c+r_{\cdotv} -2)}{ n_{\cdotv k}+c+r_{\cdotv} -1}}, \label{eqn:varmeanBNBP}
\eeq
which depends on both $c$ and $r_{\cdotv}$ that are random.
Comparing \eqref{eqn:varmeanNBP} and \eqref{eqn:varmeanBNBP}, it is clear that since $\frac{c+r_{\cdotv} -2}{n_{\cdotv k}+c+r_{\cdotv} -1}\hspace{-.3mm}\le1$ and $\frac{n_{\cdotv k}(c+r_{\cdotv} -2)}{ n_{\cdotv k}+c+r_{\cdotv} -1}<n_{\cdotv k}$ for $c+r_{\cdotv} >2$, similar to the GNBP, the BNBP can also model much more over-dispersed counts than the scaled NBP.

The variance-mean relationships expressed by (\ref{eqn:varmeanNBP}), (\ref{eqn:varmeanGNBP}), and (\ref{eqn:varmeanBNBP}) show that the GNBP and BNBP can model much more over-dispersed counts than the (scaled) NBP, and as shown in Figure 1 of \citet{NBP_CountMatrix}, given the same expected total count, while the counts in NBP random count matrices usually have small dynamic ranges, the counts in both the GNBP and BNBP matrices can contain values that are significantly above the average. In RNA-Seq, it is common to have large dynamical range for highly over-dispersed gene counts, which are likely to be better modeled by both the GNBP and BNBP than by the (scaled) NBP, as confirmed by our experiments in Section \ref{sec:3}.

\subsubsection{Inference for the BNBP}

Letting $\gamma_0 \sim \text{Gamma}(e_0,1/f_0)$, $p_j \sim \text{Beta}(a_0,b_0)$, and $c \sim \text{Gamma}(c_0,1/d_0)$, as in \citet{NBP_CountMatrix}, a Gibbs sampling iteration for the BNBP proceeds as 
\begin{align}\label{eq:BNBP_sampling}
&(\gamma_0\given -)\sim\mbox{Gamma}\bigg(e_0+K_J,\frac{1}{f_0+\psi(c+r_{\cdotv})-\psi(c)}\bigg),\notag\\
&(p_k\given-)\sim \mbox{Beta}(n_{\cdotv k},c+r_{\cdotv}),~(p_*\given-)\sim\mbox{logBeta}(\gamma_0 ,c+r_{\cdotv}), \notag\\ 
&(l_{jk}| - ) \sim\mbox{CRT}(n_{jk},r_j), 
\notag\\
&(r_j\given-)\sim\mbox{Gamma}\bigg(a_0 + l_{j\cdotv} ,\frac{1}{b_0+ p_*-\sum_{k=1}^{K_J}\ln(1-p_k)} \bigg).
\end{align}
Inside each Gibbs sampling iteration, as in \citet{NBP_CountMatrix}, an independence chain Metropolis-Hastings sampling step can be used to update the concentration parameter~$c$.

\subsubsection{BNBP-Seq differential expression analysis}

In the BNBP, since in the prior we have 
\begin{align}\E[n_{jk}\given r_j,B] &= r_j \frac{p_k}{1-p_k},~~
\mbox{var}[n_{jk}\given r_j,B] = r_j \frac{p_k}{(1-p_k)^2} = {(1-p_k)^{-1}}\E[n_{jk}\given r_j,B], \end{align}
and in the conditional posterior, if $c+r_{\cdotv}>1$, we have
\beqs\label{eq:E_BNBP}
\E[p_k/(1-p_k)\given-] = n_{\cdotv k}/(c+r_{\cdotv}-1),~~
\E[r_j\given-]=\frac{a_0 + l_{j\cdotv}}{b_0+ p_*-\sum_{k=1}^{K_J}\ln(1-p_k)}.
\eeqs
Thus one may consider that the NB sample-specific shape parameter $r_j$ accounts for the sequencing depth of sample $j$, and may compare the posterior distributions of $p_k/(1-p_k)$ to evaluate differential expression of gene $k$ between different groups. The posterior expectation of $r_j$ in the BNBP is related to the NB probability parameters of all genes, which themselves are related to $r_{\cdotv}$, the aggregation of the sample-specific scaling factors across all $J$ samples. Thus the BNBP borrows statistical strength across all the genes and samples to infer the posterior distribution of $p_k/(1-p_k)$. 
Note that for an unexpressed gene, whose total count across all samples in a group is 0, the posterior values of its $p_k$ would be fixed at 0.

Comparing \eqref{eq:E_NBP} and \eqref{eq:E_BNBP} shows that the BNBP and scaled NBP have similar gene-specific parameters, but, as in \eqref{eq:BNBP_sampling}, since $\E[l_{jk} \given -] \approx r_j\ln(n_{jk}+r_j)$ for large $n_{jk}$, for some genes whose expressions $n_{jk}$ are significantly larger than the mean expression levels, the posteriors of the sample-specific parameters $r_j$ in the BNBP also would be impacted much less than these of the sample-specific parameters $q_j$ in the scaled NBP.

\subsection{Distance between posterior distributions}

In order to compare the posterior distributions, we use the symmetric Kullback-Leibler (KL) divergence defined between two discrete distributions $P$ and $Q$ as $$KL(P,Q)=\sum \nolimits_{x} \big[p(x)-q(x)\big]\log \big[p(x)/q(x)\big].$$

Supposing $r$ is the parameter to be compared between two different groups, we estimate the symmetric KL-divergence between the posterior distributions of $r^{(1)}$ and $r^{(2)}$, the values of $r$ of the first and second groups, respectively, using collected MCMC samples. 
We first find both the minimum and maximum values of the MCMC samples of $r$ across both groups to define an interval for $r$. 
After adjusting the lower- and upper-limits of the interval as $[\max(0, Q_1-1.5*Q_{\triangle}),~~ Q_3+1.5*Q_{\triangle}]$, where $Q_1$ and $Q_3$ are 25\% and 75\% quantiles and $Q_{\triangle}=Q_3-Q_1$, we equally divide the adjusted interval into $N=100$ bins. 
For each group, we count the number of MCMC samples falling into each bin, and then normalize these bin counts to a 100 dimensional discrete probability vector, referred to as $\piv^{(1)}$ and $\piv^{(2)}$ for the first and second groups, respectively. 
 Finally, with a small constant set as $\epsilon=10^{-10}$, we calculate the symmetric KL-divergence as
\begin{equation}\label{eq:KL}
KL(\piv^{(1)}, \piv^{(2)}) = \sum_{i=1}^{N} \big(\pi_i^{(1)}-\pi_i^{(2)}\big)\log \Big( \frac{\pi_i^{(1)}+\epsilon}{\pi_i^{(2)}+\epsilon} \Big).
\end{equation}

\section{Experimental Results}
\label{sec:3}

To evaluate the proposed BNP differential expression analysis algorithms, we compare their performance on
both synthetic and real-world benchmark RNA-Seq data with those of edgeR \citep{edger}, DESeq \citep{deseq}, and baySeq \citep{bayseq}, three widely used algorithms in biomedical studies. We also present a case study on \emph{clear cell renal cell carcinoma} (ccRCC) \citep{cancer2012comprehensive}, explaining the biomedical implications obtained by differential expression analysis using both our GNBP and BNBP methods. We first consider synthetic RNA-Seq data generated under different models, 
and we 
show that the proposed GNBP and BNBP differential expression analysis algorithms consistently provide outstanding performance. 
We then consider the real-world benchmark RNA-Seq data 
extracted from the SEquencing Quality Control (SEQC) project \citep{xu2013cross,seqc} and the ccRCC case study extracted from The Cancer Genome Atlas (TCGA) \citep{mclendon2008comprehensive}. We consider the RNA-Seq data from both Beijing Genomics Institute (BGI) and the Pennsylvania State University (PSU)
provided in the SEQC project \citep{xu2013cross,seqc}, available in the R package SEQC on Bioconductor \citep{bioconductor}. 
Both the BGI and PSU datasets, which are the transcriptomic expression measurements of the RNA samples prepared at the same biological conditions but sequenced at different sequencing sites, contain the counts for approximately 26,000 genes. In our experiments, we employ sample groups A and B, which are derived from Agilent's Universal Human Reference RNA  and Life Technologies' Human Brain Reference RNA cell lines, respectively. 
We collect the counts from the first flow cells of the sequencing machines on five replicates for each group (condition).

On both synthetic and real-world RNA-Seq count data,  comparison of both the area under the receiver operating characteristic (ROC) curve (AUC-ROC) and area under the precision-recall (PR) curve (AUC-PR) shows that the proposed GNBP and BNBP algorithms clearly outperform the (scaled) NBP and previously proposed differential expression analysis algorithms, as described below in detail.

\vspace{-3mm}
\subsection{Synthetic data}
We first generate synthetic RNA-Seq data with the GNBP generative model, the BNBP generative model, or the NB distribution based procedure adopted in baySeq \citep{bayseq}. For each setting, to make the synthetic data closely resemble real-world RNA-Seq data, 
we first infer the parameters of the corresponding model on the BGI or PSU datasets from SEQC, and then generate synthetic sequencing counts using these inferred model parameters.  
To simulate samples from two different groups (conditions), each of which has 10,000 genes in five replicates, we randomly select 10\% of the genes and set them to be  
differentially expressed between the two groups, with the fold change of differentially expressed genes chosen as an adjustable parameter. For quality control, 
we discard the bottom 10\% of genes with low expressions across groups in data generation. In order to produce both up- and down-regulated differentially expressed genes, each differentially expressed gene is randomly set to be either up- or down-regulated. Below we denote $b>1$ as the fold change to be set. We use the PSU dataset for the baySeq setting and the BGI dataset for both the GNBP and BNBP settings. Using different datasets to infer model parameters and different models to generate synthetic datasets allows us to assess the robustness of various methods in different practical settings.

In the GNBP setting, if gene $k$ is up-regulated, then we generate its counts using $ \mbox{NB}(r_k,p_j)$ and $ \mbox{NB}(b\,r_k,p_j)$ for the samples in the first and second groups, respectively; whereas if gene $k$ is down-regulated, then we generate its counts using $ \mbox{NB}(r_k,p_j)$ and $ \mbox{NB}(r_k/b,p_j)$ for the five samples in the first and second groups, respectively.

In the BNBP setting, if gene $k$ is up-regulated, then we generate its counts using $ \mbox{NB}(r_j,p_k)$ and $ \mbox{NB}(r_j,p_k')$, where $p'_k$ is selected to satisfy $b p_k/(1-p_k)= p'_k/(1-p'_k)$, for the samples in the first and second groups, respectively; whereas if gene $k$ is down-regulated, then we generate its counts using $ \mbox{NB}(r_j,\tilde p_k)$ and $ \mbox{NB}(r_j,p_k)$, where $\tilde p_k$ is selected to satisfy $p_k/(1-p_k)= b \tilde p_k/(1-\tilde p_k)$, for the five samples in the first and second groups, respectively.

In the baySeq setting of \citet{bayseq} that generates a count from a NB distribution given its mean and dispersion parameters, if gene $k$ is up-regulated, then we generate its counts using $\mu_k$ and $b \mu_k$ as the means for the first and second groups, respectively; whereas if gene $k$ is down-regulated, then we generate its counts using $ \mu_k$ and $\mu_k/b$ as the means for the first and second groups, respectively.

We infer the model parameters via Gibbs sampling for the proposed BNP differential expression analysis algorithms. 
For each algorithm, we collect 1,000 MCMC samples after 1,000 burn-in iterations. The example MCMC sample trace plots in Figure~\ref{fig:trace-plot} of the Appendix suggest that the Markov chains for both the GNBP and BNBP methods converge fast and mix well, supporting the practice of performing downstream analysis with 2,000 MCMC iterations. 
For the analysis of the real-world dataset BGI on a single cluster node with Intel Xeon 2.5GHz E5-2670 v2 processor, it took around two hours for both the GNBP and BNBP methods with 2,000 MCMC iterations, about ten minutes for the other methods, including the NBP. Note that parallelization could further speed up the inference. We use the collected MCMC samples to calculate the symmetric KL divergence, as in \eqref{eq:KL}, between two groups for each gene, and rank the 
genes according to these values. 
For edgeR and DESeq, we follow the standard analysis pipelines and rank the genes using the computed $p$-values; and for baySeq, we rank the genes using model likelihoods. We set the fold change $b$ as $1.4$, $1.6$, $1.8$, or $2$ in simulating synthetic data to assess how sensitive the algorithms under study are to different levels of differential expression. For each fold change, we report the results of each algorithm based on ten independent random trials. 

\begin{figure}[!th]
	\centering
	\begin{subfigure}[b]{0.84\textwidth}
		\includegraphics[width=1.1\textwidth]{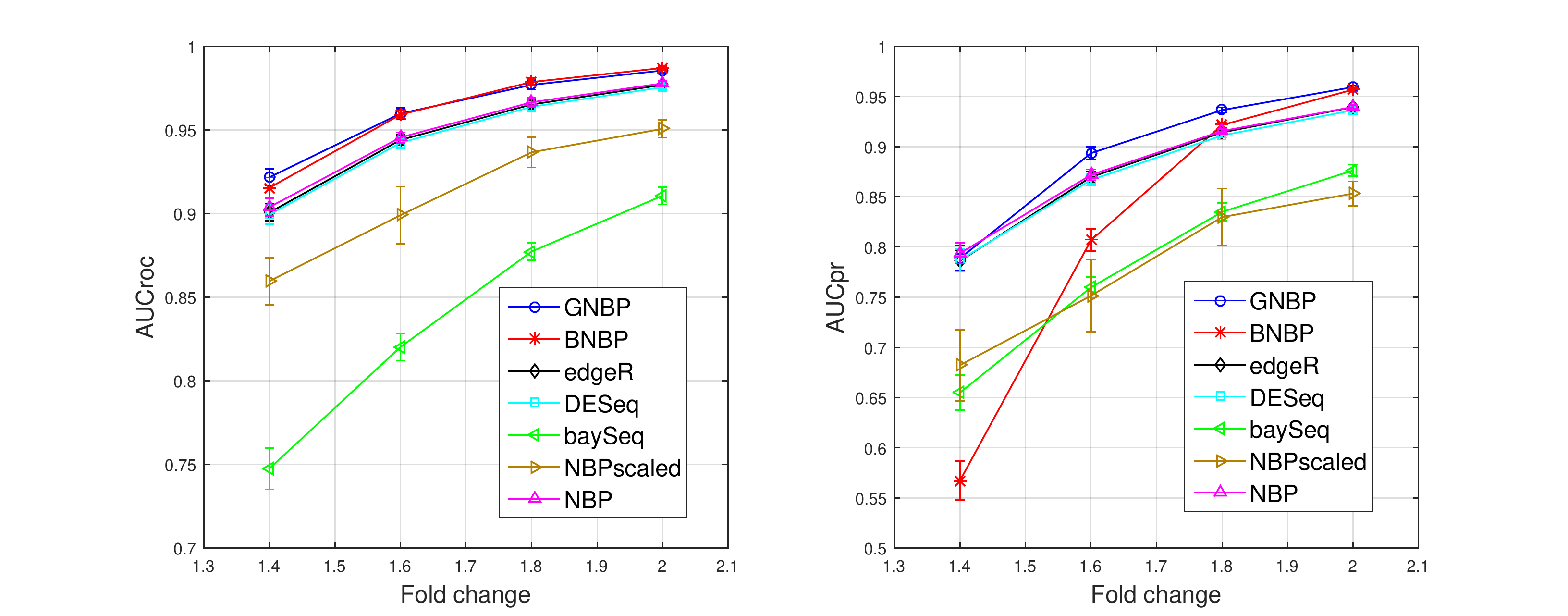}
		\caption{GNBP synthetic data}
	\end{subfigure}
	\begin{subfigure}[b]{0.84\textwidth}
		\includegraphics[width=1.1\textwidth]{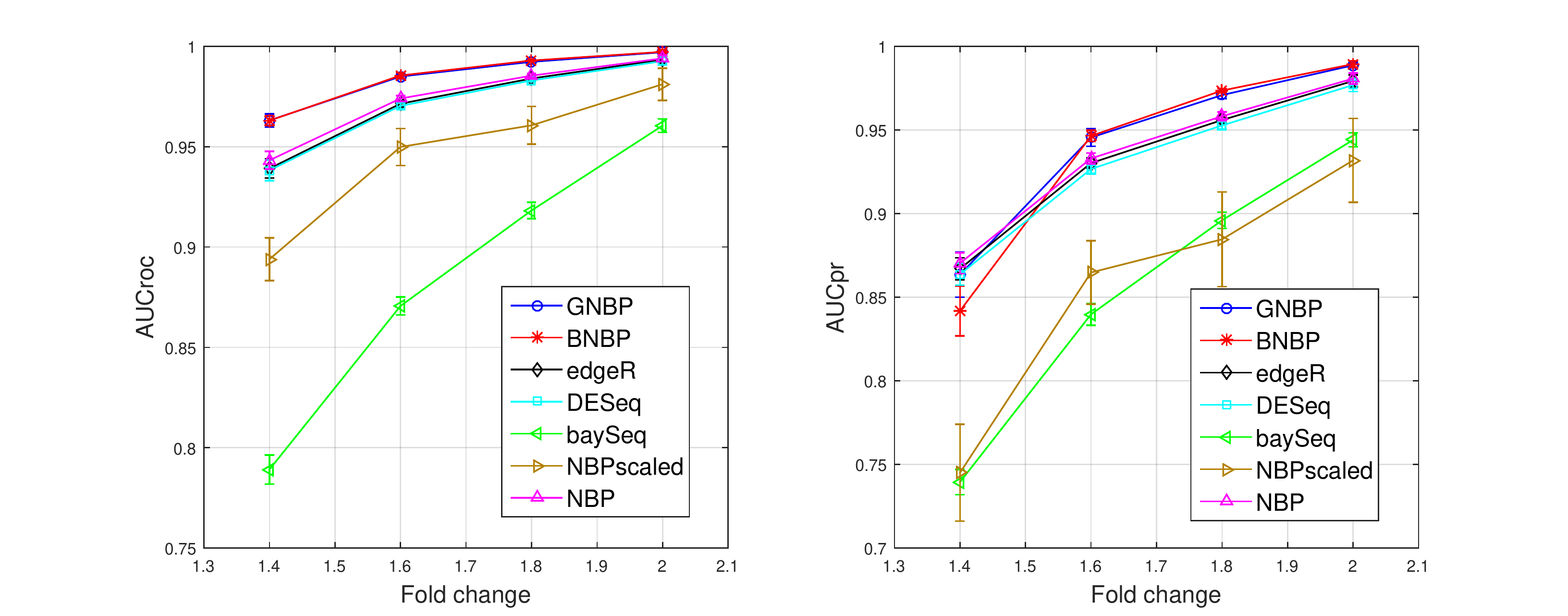}
		\caption{BNBP synthetic data}
	\end{subfigure}
	\begin{subfigure}[b]{0.84\textwidth}
		\includegraphics[width=1.1\textwidth]{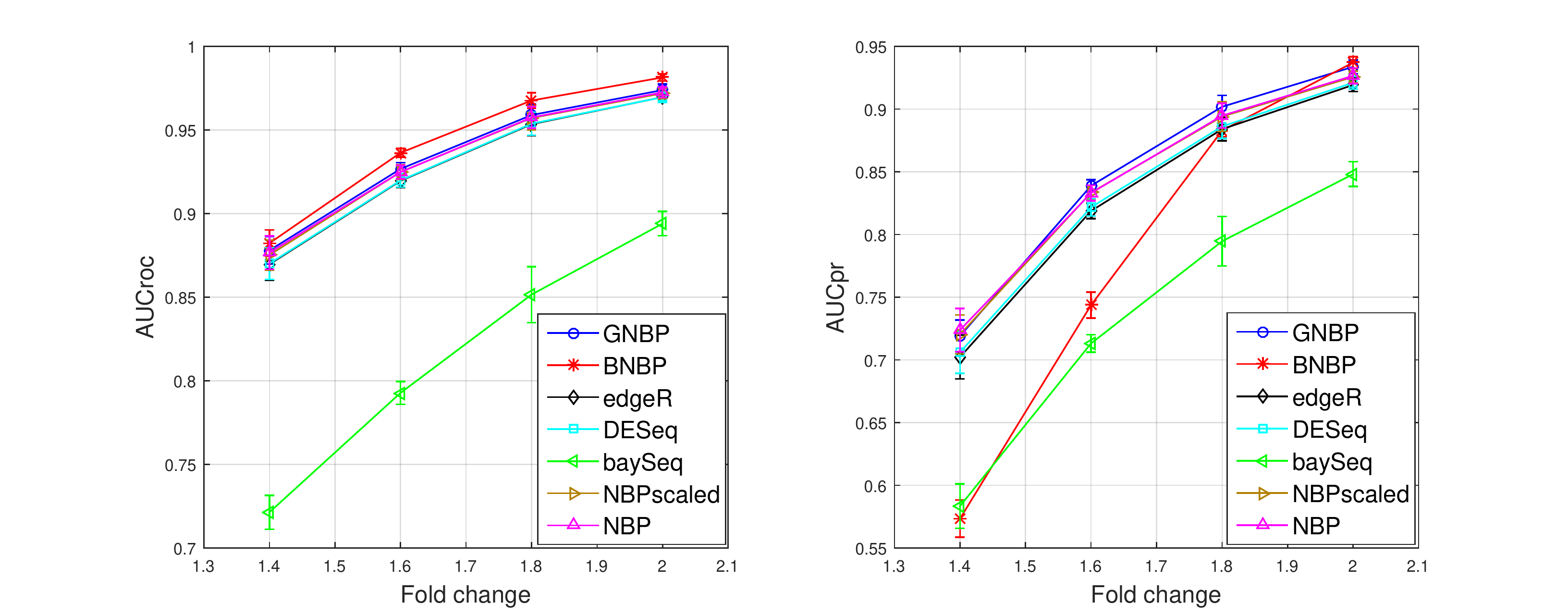}
		\caption{baySeq synthetic data}
	\end{subfigure}
	\caption{\textbf{left column}: \small AUC-ROC values, \textbf{right column}: AUC-PR values. Performance comparison of different methods  in detecting differentially expressed genes under various fold changes, using synthetic data generated under three different negative binomial distribution based models. 	}
	\label{fig:synthetic}
\end{figure}

For these three different types of synthetic data, as shown in Figure \ref{fig:synthetic}, measured by both AUC-ROC and AUC-PR, baySeq has the worst overall performance even when the synthetic data are generated based on its model assumption, followed by the scaled NBP; the NBP, DESeq, and edgeR all have similar performance; and the GNBP and BNBP clearly outperform all the other differential expression analysis algorithms. To further compare the operating characteristics of different algorithms, we show in Figure \ref{fig:simulated} in the Appendix the full ROC and PR curves for the fold change of $b=1.8$. 
We provide in the Appendix the detailed numerical values used to plot Figure \ref{fig:synthetic}, as shown in
Tables 1-6, where for each setting
the best result 
and the ones that are less than one standard deviation away from the best are highlighted in bold.

More carefully examining Figures~\ref{fig:synthetic} and~\ref{fig:simulated}, 
it is interesting to notice that for the synthetic data generated with either the GNBP or BNBP, the scaled NBP, which extends the original NBP with sample-specific scaling parameters $q_j$ to model sample sequencing depth variations, in fact clearly underperforms the original NBP. Suggesting that explicitly modeling the sample sequencing depths, using the gamma-Poisson construction of the scaled NBP, is insufficient to model the over-dispersed gene counts generated using the gamma- or beta-NB constructions. 

While the original NBP fixes $q_j=1$ and hence does not explicitly model the sample sequencing depth variations, it performs as well as both DESeq and edgeR in all three settings, which may be explained by the fact that it normalizes the posterior Poisson rates before applying them to compare the gene expression levels between two groups, a post-processing step that plays a similar role as the pre-processing normalization steps used in both DESeq and edgeR to account for different sequencing depths. 

It is also interesting to notice that while the GNBP consistently ranks the best or very close to the best, in terms of both AUC-ROC and AUC-PR, the BNBP does so only in terms of AUC-ROC. For synthetic data generated using the GNBP and baySeq, the performance of the BNBP in terms of AUC-PR quickly deteriorates as the fold change reduces from 1.8 to 1.4, suggesting a large number of false positives among the top ranked genes of the BNBP when the fold change is not sufficiently large for the GNBP synthetic data. The disparity between the performance measured by AUC-ROC and that measured by AUC-PR, which only happens for the BNBP, indicates that the BNBP employs a distinct mechanism to detect differentially expressed genes, as carefully discussed below. 

To compare the expression levels of the $k$th gene between two groups, 
the GNBP compares the posterior NB shape parameters $r_k$, whereas the BNBP compares the posterior NB probability parameters $p_k$.  One may consider that the expression level of gene $k$ is assumed to roughly follow a smooth function of the shape parameter $r_k$ in the GNBP, and a smooth function of $p_k/(1-p_k)$ in the BNBP.
 The difference between the posterior NB shape parameters $r_k$ explains the differences between both the means and dispersions, but does not explain that of the variance-to-mean ratios~(VMR), of the counts of gene~$k$ at different groups, since if $n_{jk}\sim\mbox{NB}(r_k,p_j)$, then 
$\E[n_{jk}] = r_kp_j/(1-p_j), ~\mbox{var}[n_{jk}] = \E[n_{jk}] + (\E[n_{jk}])^2/r_k$, and $\mbox{VMR}[n_{jk}] = 1+ \E[n_{jk}]/r_k$; whereas the difference between the posterior NB probability parameters $p_k$ explains the differences between both the means and VMRs of the counts of gene $k$ at different groups, since if $n_{jk}\sim\mbox{NB}(r_j,p_k)$, then 
$\E[n_{jk}] = r_jp_k/(1-p_k),~\mbox{var}[n_{jk}] = \E[n_{jk}] + (\E[n_{jk}])^2/r_j$, and $\mbox{VMR}[n_{jk}] = 1/(1-p_k)$. Therefore, for the counts of a gene generated with the GNBP, if the $r_k$ is small, a small change in its value may lead to a significant change of $\mbox{VMR}[n_{jk}] = 1+ (\E[n_{jk}])/r_k$, which implies that a large difference in a gene's VMRs between two groups may not be taken by the GNBP as a strong evidence for differential expression. 
By contrast, since the gene-specific parameter $p_k$ in the BNBP is explicitly responsible for the VMR, a large difference in a gene's VMRs between two groups may encourage the BNBP to rank that gene as strongly differentially expressed, which may be used to explain why the BNBP has good AUC-ROC but poor AUC-PR if the fold change is small for the GNBP synthetic data. In practice, however, it is often unclear whether it is the change of the quadratic relationship between the variance and mean, as captured by the NB dispersion parameter, or the VMR, as captured by the NB probability parameter, that is responsible for the change of a gene's expression level. Thus it is often unclear whether the GNBP or BNBP would be a better choice for a real dataset, and it seems promising to combine the advantages of both for differential expression analysis, an attractive research topic beyond the scope of the paper that is to be investigated in our future study. 

To more comprehensively evaluate the proposed methods, we consider several additional application scenarios. The performance comparisons with baySeq synthetic data under these different scenarios are shown in Figure~\ref{fig:synthetic2} in the Appendix. 
We first assess the sensitivity of different methods to the ratio of up- and down-regulated genes among the set of truly differentially expressed genes, which, the same as before, take 10\% of the total number of genes. We assume a fold change of 2 for these truly differently expressed genes, and vary the percentage of up-regulated (down-regulated) genes among them from 20\% (80\%) to 40\% (60\%), 60\% (40\%), and 80\% (20\%).
As shown in Figure~\ref{fig:synthetic2}(a), 
while the GNBP, BNBP, edgeR, and DESeq all show robustness to the change of that percentage, the performance of both the NBP based and baySeq methods significantly deteriorates as one increases the imbalance between the numbers of up- and down-regulated genes. We also note that both the GNBP and BNBP successfully preserve their out-performance margins under various ratios of up- to down-regulated genes.
	
To examine how the performance changes with the sample size, we consider increasing the number of samples for each group from 4, to 8, 12, and 16. This is a sensible choice, since in practice the number of samples per condition is often smaller than 16. 
In this experiment, 10\% of genes are equally likely to be up- or down-regulated with a fold change of 2. Figure~\ref{fig:synthetic2}(b)  illustrates the error bar plots for both the AUC of ROC curve and that of PR curve, under different sample sizes over 10 random trials. All methods show consistent improvements as the number of replicates in each group increases, which agrees with the expectation that more samples provide more information to assist 
parameter inference. In addition, we consider 100 genes with different sample sizes to investigate the performance of the proposed methods in the setting with a large sample size but a small number of genes. Similar to previous simulations, 10\% of the genes are assumed to be differentially expressed with a fold change of 2, and the number of replicates in each group is increased from 4 to 6, 8, 10, 20, 40, 60, 80, and 100. Figure~\ref{fig:small} shows the error bar plots for both the AUC of ROC curve and that of PR curve, under different sample sizes over 10 random trials. As expected, adding more samples consistently enhances the recovery of true differential expression state of the genes for all methods, and when the number of samples reaches 100, almost all methods perform perfectly.

Last but not least, Figure~\ref{fig:synthetic2}(c) shows the box plots of the AUCs of ROC and PR curves when the true fold change of differentially expressed genes is uniformly distributed within the interval $[1.4,2]$. The BNBP stands out as the best performing method followed by the GNBP, suggesting that the superior results of the proposed methods in previous simulations do not rely on setting the fold change to a fixed constant.

\vspace{-3mm}
\subsection{SEQC benchmark RNA-Seq data and case study}
\label{sec:verify}
In order to 
characterize various RNA-Seq technologies and quantification pipelines in the SEQC project \citep{xu2013cross,seqc}, the same RNA samples for a comprehensive group of control genes are analyzed based on quantitative Reverse Transcription Polymerase Chain Reaction (qRT-PCR) using TaqMan assays \citep{joyce2002quantitative}, which is referred as the TaqMan benchmark data \citep{maqc,maqc2}. For sample groups A and B, the expression intensity values of 955 selected control genes have been derived in the TaqMan qPT-PCR analysis for sequencing benchmarking. Without knowing in practice which genes are truly differentially expressed between different conditions, we consider thresholding the qRT-PCR expression ratios between different conditions at a certain value to define the ground-truth set of differentially expressed genes. Based on these 955 genes in the TaqMan data, we evaluate the performance of different differential expression analysis pipelines. Note that 
although the replicates in SEQC are technical, they show notable amount of over-dispersion and have been used in the literature 
as a standard benchmark for assessing differential expression analysis tools \citep{comprehensive}.

\begin{figure}[!t] 
	\centering
	\begin{subfigure}[b]{.83\textwidth}
		\includegraphics[width=1\textwidth]{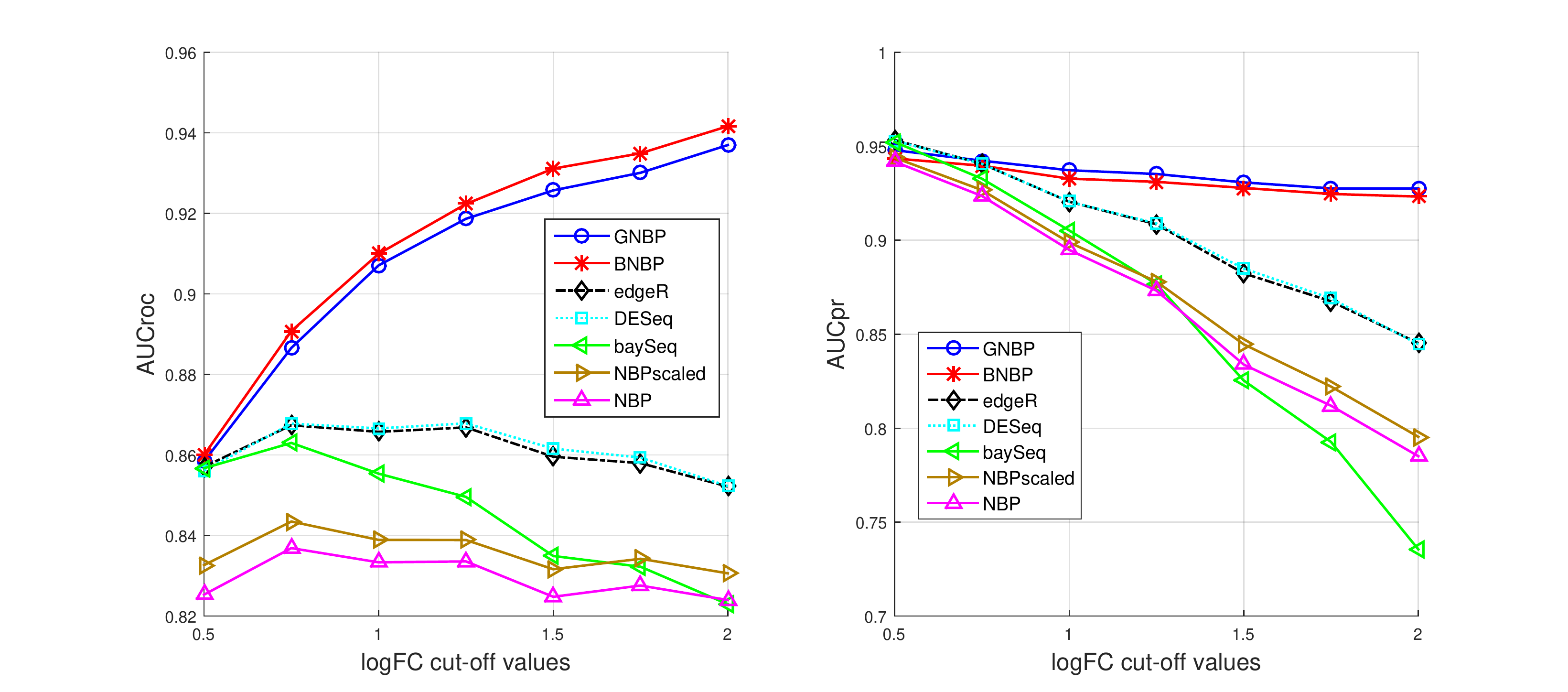}
		\caption{BGI dataset}
	\end{subfigure}
	
	\begin{subfigure}[b]{0.83\textwidth}
		\includegraphics[width=1\textwidth]{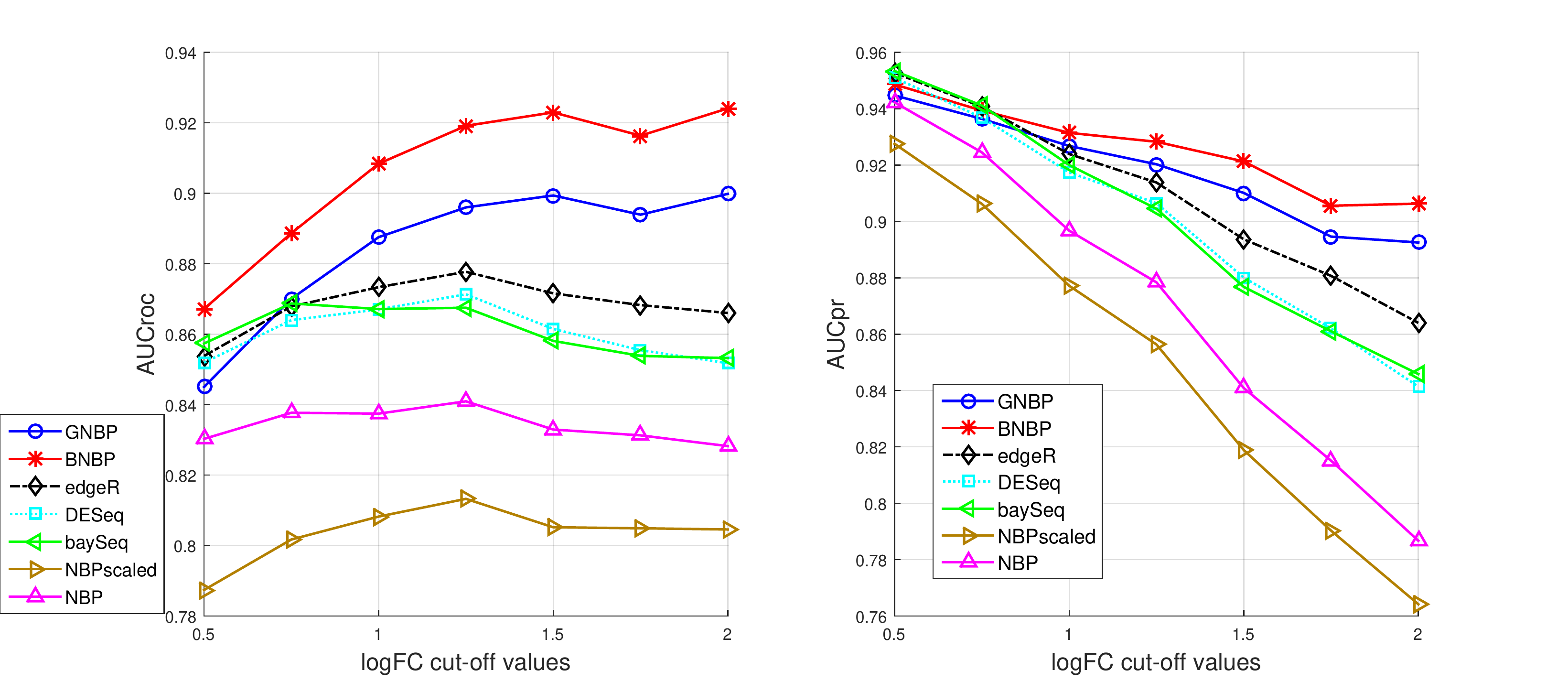}
		\caption{PSU dataset}
	\end{subfigure}
	\caption{\textbf{left column}: AUC-ROC values, \textbf{right column}: AUC-PR values. Performance comparison of different methods in detecting differentially expressed genes on real-world benchmark RNA-Seq data from the SEQC project.} 
	\label{fig:taqman}
\end{figure}

While it is unknown which genes are truly differentially expressed for both the BGI and PSU RNA-Seq data, we rely on the qRT-PCR expression intensity of the 955 genes in the TaqMan data and set different cut-offs for the binary logarithm (log2) of the qRT-PCR expression ratio to define ``truly'' differentially expressed genes. We increase this log2 cut-off value gradually from 0.5 to 2, and calculate both AUC-ROC and AUC-PR. The symmetric KL divergence is used to assess differential expression. 
As shown in Figure \ref{fig:taqman} for both the BGI and PSU datasets, the GNBP and BNBP outperform all the other methods in both ROC and PR analyses with significant margins. Note that the performance gains of the GNBP and BNBP over the other methods become more significant as one increases the log2 cut-off for the qRT-PCR expression ratio, which reduces the number of genes that are considered as truly differentially expressed.

Comparing Figure \ref{fig:synthetic} for synthetic data with Figure \ref{fig:taqman} for real-world data, one may notice that while both the AUC-ROC and AUC-PR curves in Figure \ref{fig:synthetic} seem to monotonically increase as the fold change increases, the AUC-ROC and AUC-PR curves in Figure~\ref{fig:taqman} do not necessarily share similar trends. It is not surprising to observe these seemingly distinct behaviors, since for the synthetic data in Figure \ref{fig:synthetic}, the set of truly differentially expressed genes are fixed and known exactly, remaining unchanged regardless of how one sets the fold change that is used to detect differentially expressed genes, whereas for the real-world data in Figure \ref{fig:taqman}, the number of genes considered as truly expressed reduces as the cut-off value of the qRT-PCR expression ratio increases. In addition, we note that the results of edgeR, DESeq, and baySeq on both the BGI and PSU real datasets reported in this paper are similar to those reported in \citet{comprehensive}.

\begin{figure}[!t] 
	\centering
	\includegraphics[width=0.86\textwidth]{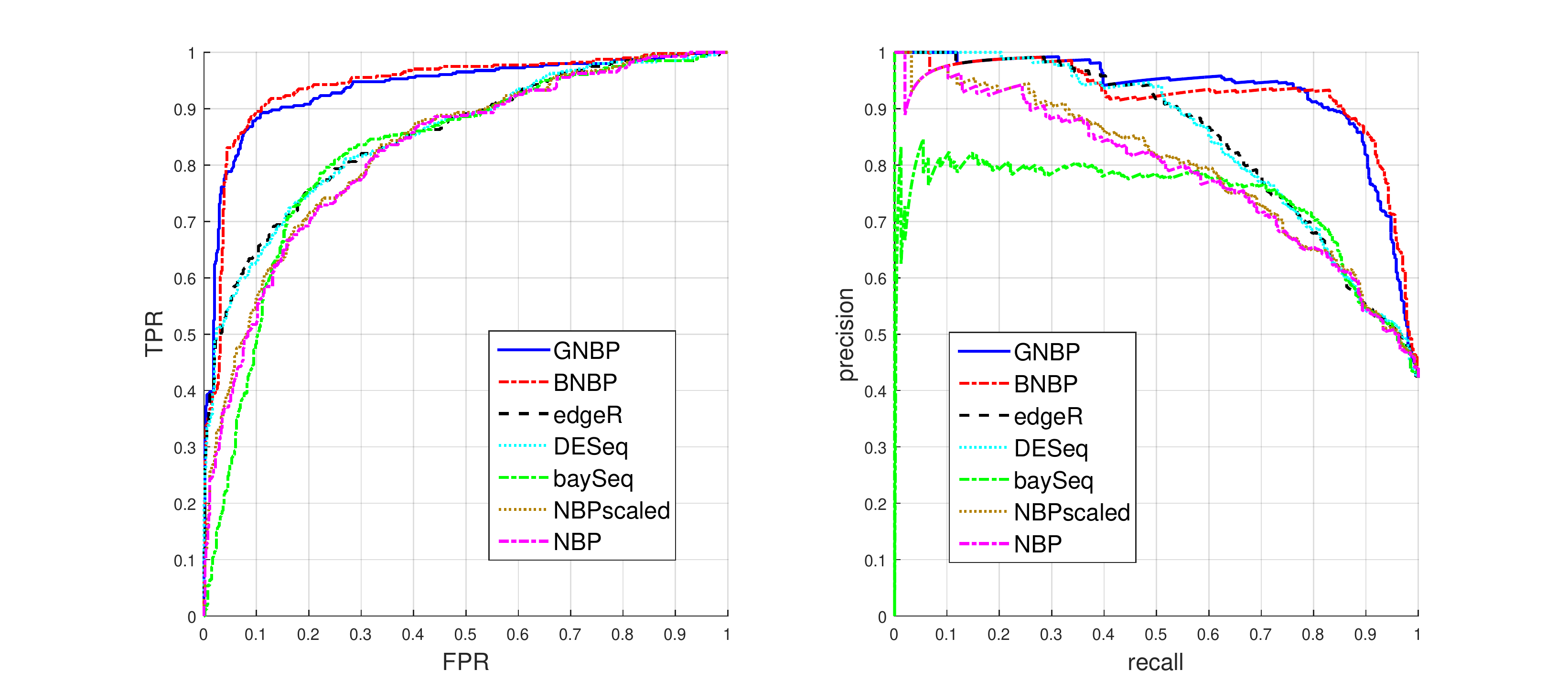}
	\caption{\textbf{left}: ROC curves, \textbf{right}: Precision-Recall (PR) curves. Performance comparison of different methods  with the log2 cut-off value fixed at 2 for the BGI dataset from the SEQC project.}
	\label{fig:roc}
\end{figure}

To investigate the experimental results more thoroughly, we fix the true positives and negatives at the log2 cut-off value of 2 and illustrate the ROC and PR curves for BGI dataset in \figurename{~\ref{fig:roc}}. In addition, we show in  Table~\ref{table:AUC0.1} the
area under the ROC curve for the range with $FPR \leq 0.1$ and area under the PR curve for the range with $\mbox{Recall} \leq 0.1$ for various algorithms.  It is clear  that both the GNBP and BNBP not only have higher AUC-ROC and AUC-PR, but also outperform all the other methods in almost all regions of the ROC and PR curves.

In addition to showing the ROC and PR curves, we also plot the number of false discoveries to highlight the performance on the top ranked genes. 
Since there are 400 truly differentially expressed genes based on the log2 cut-off value of 2, the top 400 genes detected by each approach are selected and the number of false discoveries are plotted. It is clear from Figure \ref{fig:fd} in the Appendix that both the GNBP and BNBP return much smaller number of false positives in comparison to all the other differential expression analysis algorithms.

\subsection{Case study: clear cell renal cell carcinoma}

For further illustration,  we provide a case study with both the GNBP and BNBP methods on clear cell renal cell carcinoma (ccRCC),   the most common type of kidney cancer that is closely related to genes involved in regulating cellular metabolism \citep{cancer2012comprehensive}. 
We collect 21 samples from The Cancer Genome Atlas (TCGA) \citep{mclendon2008comprehensive}, 11 of which are from the patients diagnosed with ccRCC and kidney as the primary site before any treatment procedures and the remaining 10 samples are from normal persons. These samples contain roughly 1.2 billion reads mapped to one of the 60,843 genomic locations annotated by the Ensembl database. The raw counts produced by HTSeq \citep{anders2014htseq} are downloaded from the TCGA Website (see https://gdc-portal.nci.nih.gov/). 
We first filter out the genes whose total counts over all samples are less than 20. We apply both the GNBP and BNBP methods to detect differentially expressed genes, where after 1,000 burn-in iterations, 1,000 posterior samples are collected to calculate the KL-divergence between two different conditions to rank the genes. 
Similar to EdgeR and DESeq, our BNP based methods provide new statistical tools for differential expression analysis of RNA-Seq data and can help identify critical biomarkers in biomedical applications.

The top differentially expressed gene identified by the BNBP method is Interleukin-32 (IL-32), a proinflammatory cytokine that acts as a significant pathogenetic factor in various diseases and malignancies. It is reported as a potential prognostic factor for predicting outcomes in patients with ccRCC \citep{lee2012overexpression}. The third gene in the ranking list is FHL1, for which clinical analyses suggest its expression is suppressed in some human tumors, including those of the breast, kidney, and prostate and it suppresses metastatic cell growth \citep{li2008coordinate}. The fourth gene, VIM, is another potential oncogene regulated by tumor suppressive microRNA-138 \citep{yamasaki2012tumor}. The fifth gene, Progranulin (PGRN), is a pluripotent secreted growth factor that mediates cell cycle progression and cell motility, and is highly expressed in aggressive cancer cell lines and clinical specimens, including breast, ovarian, and renal cancers as well as gliomas \citep{he2003progranulin}. The ninth gene, SLC12A1, is a tumor suppressor that is reported to lose its function in ccRCC as a result of increased levels of some microRNAs such as miR-142-3p and miR-185 \citep{liu2010identifying}. The tenth gene, PGK1, is a target gene of MYC pathway, which potentially has an essential role in the proliferation of ccRCC cells \citep{tang2009myc}.

The top differentially expressed gene identified by the GNBP method is UMOD, encoding uromodulin, the major protein secreted in the normal urine that has a link to kidney chronic disease. It has been considered as a potential therapeutic target to preserve renal function \citep{trudu2013common}. The fourth gene, MT-ND4, is a mitochondrial gene whose mutation leads to complex I enzyme deficiency found in renal oncocytoma \citep{gasparre2008clonal}. The seventh gene, AQP2, involved in regulating the homeostasis of water-electrolyte balances has been found to be the most significantly down-regulated in ccRCCs \citep{zhou2010integrated}. The ninth gene, COX-2 is associated with several clinicopathological factors, and is conjectured to play an important role in tumor cell proliferation and MMP-2 expression \citep{miyata2003expression}.

In addition to these top ranked genes, several other important genes, previously shown to have strong connections to renal cell carcinoma, are also ranked in the top 5\%  by both the GNBP and BNBP methods. In the following we list some {well-studied ccRCC biomarkers or genetic risk factors in the literature}.

\begin{itemize}
		\item 
	The CA9 gene, whose corresponding encoded protein is a tumor-associated antigen, is well known in the literature as a reliable diagnostic biomarker of clear cell carcinoma of kidney \citep{liao1997identification,bui2003carbonic}.
	
	\item The MT1G gene 
	has been demonstrated to have frequent occurrence of Methylation of CpG dinucleotides in the promoter region, which is a major mechanism of tumor suppressor genes inactivation in renal cell carcinoma \citep{morris2003multigene}.
		
	\item 
	A member of the AP-2 family of transcription factors, TFAP2B, 
	functions as both a transcriptional activator and repressor, required for proper terminal differentiation and function of renal tubular epithelia for the survival of renal epithelial cells during renal development \citep{tun2010pathway}. Its down-regulation in ccRCC has been verified by Microarray profiling and PCR \citep{tun2010pathway}. 
	
	\item The DACH1 gene is a molecular marker of renal cell carcinoma where its expression remarkably decreases and the restoration of DACH1 function in renal clear cell cancer cells inhibits \emph{in vitro} cellular proliferation, S phase progression, clone formation, and \emph{in vivo} tumor growth \citep{chu2014dach1}.	
		
	\item ANGPT2 is a member of the angiopoietin family, which plays a pivotal role in angiogenesis during cancer development and metastasis \citep{lu2014mir}. Upregulation of ANGPT2 has been reported in various human cancers including renal cell carcinoma \citep{bullock2010plasma,baldewijns2007high}. \end{itemize}

\vspace{-4mm}
\section{Conclusions}
\vspace{-3mm}
\label{sec:conc}
We exploit Bayesian nonparametric priors, including the gamma-Poisson, gamma-negative binomial, and beta-negative binomial processes, to model RNA sequencing count matrices. With different sequencing depths captured by sample-specific model parameters, the posterior distributions of certain gene-specific model parameters are used to detect the genes that are differentially expressed between different conditions. With the model parameters inferred by borrowing statistical strength across both the genes and samples, there is no need to adjust the raw counts using heuristics before downstream analyses, an important pre-processing step that is often required in previously proposed algorithms. Example results on both synthetic and real-world RNA-Seq data demonstrate the state-of-the-art performance of both the gamma- and beta-negative binomial processes based differential expression analysis algorithms. 
Given the success of the proposed random-process-based algorithms in differential expression analysis, it is of interest to investigate Bayesian nonparametric algorithms for many other real-world applications in biomedicine that require analyzing next-generation sequencing data. 



\section{Acknowledgments}
We thank the editor and three reviewers for insightful and constructive comments and suggestions. 
We thank Texas A\&M High Performance Research Computing 
and Texas Advanced Computing Center for providing computational resources to perform experiments in this paper. X. Qian acknowledges the support of CAREER award 1553281 from the U.S. National Science
Foundation.


\begin{spacing}{1}
\small
\bibliographystyle{abbrvnat} 

\bibliography{Bibliography-MM-MC,References052016,reference}

\begin{thebibliography}{67}
\providecommand{\natexlab}[1]{#1}
\providecommand{\url}[1]{\texttt{#1}}
\expandafter\ifx\csname urlstyle\endcsname\relax
  \providecommand{\doi}[1]{doi: #1}\else
  \providecommand{\doi}{doi: \begingroup \urlstyle{rm}\Url}\fi

\bibitem[Anders and Huber(2010)]{deseq}
S.~Anders and W.~Huber.
\newblock Differential expression analysis for sequence count data.
\newblock \emph{Genome Biology}, 11\penalty0 (10):\penalty0 R106, 2010.

\bibitem[Anders et~al.(2014)Anders, Pyl, and Huber]{anders2014htseq}
S.~Anders, P.~T. Pyl, and W.~Huber.
\newblock Htseq--a python framework to work with high-throughput sequencing
  data.
\newblock \emph{Bioinformatics}, page btu638, 2014.

\bibitem[Baldewijns et~al.(2007)Baldewijns, Thijssen, Van~den Eynden,
  Van~Laere, Bluekens, Roskams, Van~Poppel, De~Bruine, Griffioen, and
  Vermeulen]{baldewijns2007high}
M.~Baldewijns, V.~Thijssen, G.~Van~den Eynden, S.~Van~Laere, A.~Bluekens,
  T.~Roskams, H.~Van~Poppel, A.~De~Bruine, A.~Griffioen, and P.~Vermeulen.
\newblock High-grade clear cell renal cell carcinoma has a higher angiogenic
  activity than low-grade renal cell carcinoma based on histomorphological
  quantification and qrt--pcr mrna expression profile.
\newblock \emph{British Journal of Cancer}, 96\penalty0 (12):\penalty0
  1888--1895, 2007.

\bibitem[Bliss and Fisher(1953)]{NB_Fitting_53}
C.~I. Bliss and R.~A. Fisher.
\newblock Fitting the negative binomial distribution to biological data.
\newblock \emph{Biometrics}, 9\penalty0 (2):\penalty0 176--200, 1953.

\bibitem[Broderick et~al.(2015)Broderick, Mackey, Paisley, and
  Jordan]{NBPJordan}
T.~Broderick, L.~Mackey, J.~Paisley, and M.~I. Jordan.
\newblock Combinatorial clustering and the beta negative binomial process.
\newblock \emph{IEEE Trans. Pattern Anal. Mach. Intell.}, 2015.

\bibitem[Bui et~al.(2003)Bui, Seligson, Han, Pantuck, Dorey, Huang, Horvath,
  Leibovich, Chopra, Liao, et~al.]{bui2003carbonic}
M.~H. Bui, D.~Seligson, K.-r. Han, A.~J. Pantuck, F.~J. Dorey, Y.~Huang,
  S.~Horvath, B.~C. Leibovich, S.~Chopra, S.-Y. Liao, et~al.
\newblock Carbonic anhydrase ix is an independent predictor of survival in
  advanced renal clear cell carcinoma implications for prognosis and therapy.
\newblock \emph{Clinical Cancer Research}, 9\penalty0 (2):\penalty0 802--811,
  2003.

\bibitem[Bullard et~al.(2010)Bullard, Purdom, Hansen, and
  Dudoit]{bullard2010evaluation}
J.~H. Bullard, E.~Purdom, K.~D. Hansen, and S.~Dudoit.
\newblock Evaluation of statistical methods for normalization and differential
  expression in {mRNA-Seq} experiments.
\newblock \emph{BMC Bioinformatics}, 11\penalty0 (1):\penalty0 94, 2010.

\bibitem[Bullock et~al.(2010)Bullock, Zhang, O'Neill, Percy, Sukhatme, Mier,
  Atkins, and Bhatt]{bullock2010plasma}
A.~Bullock, L.~Zhang, A.~O'Neill, A.~Percy, V.~Sukhatme, J.~Mier, M.~Atkins,
  and R.~Bhatt.
\newblock Plasma angiopoietin-2 (ang2) as an angiogenic biomarker in renal cell
  carcinoma (rcc).
\newblock In \emph{ASCO Annual Meeting Proceedings}, volume~28, page 4630,
  2010.

\bibitem[{{Cancer Genome Atlas Research Network}}
  et~al.(2012)]{cancer2012comprehensive}
{{Cancer Genome Atlas Research Network}} et~al.
\newblock Comprehensive molecular characterization of clear cell renal cell
  carcinoma.
\newblock \emph{Nature}, 499\penalty0 (7456):\penalty0 43--49, 2012.

\bibitem[Caron et~al.(2014)Caron, Teh, and Murphy]{CarTehMur2013a}
F.~Caron, Y.~W. Teh, and B.~T. Murphy.
\newblock {B}ayesian nonparametric {P}lackett-{L}uce models for the analysis of
  clustered ranked data.
\newblock \emph{Annal of Applied Statistics}, 2014.

\bibitem[Chu et~al.(2014)Chu, Han, Yuan, Nie, Wu, Chen, Guo, Yu, and
  Wu]{chu2014dach1}
Q.~Chu, N.~Han, X.~Yuan, X.~Nie, H.~Wu, Y.~Chen, M.~Guo, S.~Yu, and K.~Wu.
\newblock Dach1 inhibits cyclin d1 expression, cellular proliferation and tumor
  growth of renal cancer cells.
\newblock \emph{Journal of Hematology \& Oncology}, 7\penalty0 (1):\penalty0 1,
  2014.

\bibitem[Datta and Nettleton(2014)]{datta2014statistical}
S.~Datta and D.~Nettleton.
\newblock \emph{Statistical Analysis of Next Generation Sequencing Data}.
\newblock Frontiers in Probability and the Statistical Sciences. Springer
  International Publishing, 2014.

\bibitem[Dillies et~al.(2013)Dillies, Rau, Aubert, Hennequet-Antier,
  Jeanmougin, Servant, Keime, Marot, Castel, Estelle,
  et~al.]{dillies2013comprehensive}
M.-A. Dillies, A.~Rau, J.~Aubert, C.~Hennequet-Antier, M.~Jeanmougin,
  N.~Servant, C.~Keime, G.~Marot, D.~Castel, J.~Estelle, et~al.
\newblock A comprehensive evaluation of normalization methods for illumina
  high-throughput {RNA} sequencing data analysis.
\newblock \emph{Briefings in Bioinformatics}, 14\penalty0 (6):\penalty0
  671--683, 2013.

\bibitem[Ferguson(1973)]{ferguson73}
T.~S. Ferguson.
\newblock A {B}ayesian analysis of some nonparametric problems.
\newblock \emph{Ann. Statist.}, 1\penalty0 (2):\penalty0 209--230, 1973.

\bibitem[Gasparre et~al.(2008)Gasparre, Hervouet, de~Laplanche, Demont,
  Pennisi, Colombel, M{\`e}ge-Lechevallier, Scoazec, Bonora, Smeets,
  et~al.]{gasparre2008clonal}
G.~Gasparre, E.~Hervouet, E.~de~Laplanche, J.~Demont, L.~F. Pennisi,
  M.~Colombel, F.~M{\`e}ge-Lechevallier, J.-Y. Scoazec, E.~Bonora, R.~Smeets,
  et~al.
\newblock Clonal expansion of mutated mitochondrial dna is associated with
  tumor formation and complex i deficiency in the benign renal oncocytoma.
\newblock \emph{Human Molecular Genetics}, 17\penalty0 (7):\penalty0 986--995,
  2008.

\bibitem[Gentleman et~al.(2004)Gentleman, Carey, Bates, Bolstad, Dettling,
  Dudoit, Ellis, Gautier, Ge, Gentry, et~al.]{bioconductor}
R.~C. Gentleman, V.~J. Carey, D.~M. Bates, B.~Bolstad, M.~Dettling, S.~Dudoit,
  B.~Ellis, L.~Gautier, Y.~Ge, J.~Gentry, et~al.
\newblock Bioconductor: open software development for computational biology and
  bioinformatics.
\newblock \emph{Genome Biology}, 5\penalty0 (10):\penalty0 R80, 2004.

\bibitem[Greenwood and Yule(1920)]{Yule}
M.~Greenwood and G.~U. Yule.
\newblock An inquiry into the nature of frequency distributions representative
  of multiple happenings with particular reference to the occurrence of
  multiple attacks of disease or of repeated accidents.
\newblock \emph{J. R. Stat. Soc.}, 1920.

\bibitem[Hardcastle and Kelly(2010)]{bayseq}
T.~J. Hardcastle and K.~A. Kelly.
\newblock bayseq: empirical bayesian methods for identifying differential
  expression in sequence count data.
\newblock \emph{BMC Bioinformatics}, 11\penalty0 (1):\penalty0 422, 2010.

\bibitem[He and Bateman(2003)]{he2003progranulin}
Z.~He and A.~Bateman.
\newblock Progranulin (granulin-epithelin precursor, pc-cell-derived growth
  factor, acrogranin) mediates tissue repair and tumorigenesis.
\newblock \emph{Journal of Molecular Medicine}, 81\penalty0 (10):\penalty0
  600--612, 2003.

\bibitem[Hjort(1990)]{Hjort}
N.~L. Hjort.
\newblock Nonparametric {B}ayes estimators based on beta processes in models
  for life history data.
\newblock \emph{Ann. Statist.}, 1990.

\bibitem[Joyce(2002)]{joyce2002quantitative}
C.~Joyce.
\newblock Quantitative rt-pcr.
\newblock \emph{RT-PCR Protocols}, pages 83--92, 2002.

\bibitem[Kingman(1993)]{PoissonP}
J.~F.~C. Kingman.
\newblock \emph{Poisson Processes}.
\newblock Oxford University Press, 1993.

\bibitem[Kullback and Leibler(1951)]{kullback1951information}
S.~Kullback and R.~A. Leibler.
\newblock On information and sufficiency.
\newblock \emph{Annals of Mathematical Statistics}, 22\penalty0 (1):\penalty0
  79--86, 1951.

\bibitem[Lee et~al.(2012)Lee, Liang, Huang, Lim, Yoon, Lee, and
  Kim]{lee2012overexpression}
H.-J. Lee, Z.~L. Liang, S.~M. Huang, J.-S. Lim, D.-Y. Yoon, H.-J. Lee, and
  J.~M. Kim.
\newblock Overexpression of il-32 is a novel prognostic factor in patients with
  localized clear cell renal cell carcinoma.
\newblock \emph{Oncology Letters}, 3\penalty0 (2):\penalty0 490--496, 2012.

\bibitem[Li and Tibshirani(2013)]{li2013finding}
J.~Li and R.~Tibshirani.
\newblock Finding consistent patterns: a nonparametric approach for identifying
  differential expression in {RNA-Seq} data.
\newblock \emph{Statistical Methods in Medical Research}, 22\penalty0
  (5):\penalty0 519--536, 2013.

\bibitem[Li et~al.(2011)Li, Witten, Johnstone, and
  Tibshirani]{li2011normalization}
J.~Li, D.~M. Witten, I.~M. Johnstone, and R.~Tibshirani.
\newblock Normalization, testing, and false discovery rate estimation for
  {RNA}-sequencing data.
\newblock \emph{Biostatistics}, page kxr031, 2011.

\bibitem[Li et~al.(2008)Li, Jia, Shen, Ichikawa, Jarvik, Nagele, and
  Goldberg]{li2008coordinate}
X.~Li, Z.~Jia, Y.~Shen, H.~Ichikawa, J.~Jarvik, R.~G. Nagele, and G.~S.
  Goldberg.
\newblock Coordinate suppression of sdpr and fhl1 expression in tumors of the
  breast, kidney, and prostate.
\newblock \emph{Cancer Science}, 99\penalty0 (7):\penalty0 1326--1333, 2008.

\bibitem[Liao et~al.(1997)Liao, Aurelio, Jan, Zavada, and
  Stanbridge]{liao1997identification}
S.-Y. Liao, O.~N. Aurelio, K.~Jan, J.~Zavada, and E.~J. Stanbridge.
\newblock Identification of the mn/ca9 protein as a reliable diagnostic
  biomarker of clear cell carcinoma of the kidney.
\newblock \emph{Cancer Research}, 57\penalty0 (14):\penalty0 2827--2831, 1997.

\bibitem[Liu et~al.(2010)Liu, Brannon, Reddy, Alexe, Seiler, Arreola, Oza, Yao,
  Juan, Liou, et~al.]{liu2010identifying}
H.~Liu, A.~R. Brannon, A.~R. Reddy, G.~Alexe, M.~W. Seiler, A.~Arreola, J.~H.
  Oza, M.~Yao, D.~Juan, L.~S. Liou, et~al.
\newblock Identifying mrna targets of microrna dysregulated in cancer: with
  application to clear cell renal cell carcinoma.
\newblock \emph{BMC Systems Biology}, 4\penalty0 (1):\penalty0 1, 2010.

\bibitem[Lorenz et~al.(2014)Lorenz, Gill, Mitra, and Datta]{lorenz2014using}
D.~J. Lorenz, R.~S. Gill, R.~Mitra, and S.~Datta.
\newblock Using {RNA-seq} data to detect differentially expressed genes.
\newblock In \emph{Statistical Analysis of Next Generation Sequencing Data},
  pages 25--49. Springer, 2014.

\bibitem[Love et~al.(2014)Love, Huber, and Anders]{deseq2}
M.~I. Love, W.~Huber, and S.~Anders.
\newblock Moderated estimation of fold change and dispersion for {RNA-seq} data
  with {DESeq2}.
\newblock \emph{Genome Biology}, 15\penalty0 (12):\penalty0 1--21, 2014.

\bibitem[Lov{\'e}n et~al.(2012)Lov{\'e}n, Orlando, Sigova, Lin, Rahl, Burge,
  Levens, Lee, and Young]{loven2012revisiting}
J.~Lov{\'e}n, D.~A. Orlando, A.~A. Sigova, C.~Y. Lin, P.~B. Rahl, C.~B. Burge,
  D.~L. Levens, T.~I. Lee, and R.~A. Young.
\newblock Revisiting global gene expression analysis.
\newblock \emph{Cell}, 151\penalty0 (3):\penalty0 476--482, 2012.

\bibitem[Lu et~al.(2014)Lu, Ji, Li, Zhai, Zhao, Jiang, Zhang, Nie, and
  Yu]{lu2014mir}
R.~Lu, Z.~Ji, X.~Li, Q.~Zhai, C.~Zhao, Z.~Jiang, S.~Zhang, L.~Nie, and Z.~Yu.
\newblock mir-145 functions as tumor suppressor and targets two oncogenes,
  angpt2 and nedd9, in renal cell carcinoma.
\newblock \emph{Journal of Cancer Research and Clinical Oncology}, 140\penalty0
  (3):\penalty0 387--397, 2014.

\bibitem[{MAQC Consortium}(2010)]{maqc2}
{MAQC Consortium}.
\newblock The {MicroArray Quality Control {(MAQC)-II}} study of common
  practices for the development and validation of microarray-based predictive
  models.
\newblock \emph{Nature Biotechnology}, 28\penalty0 (8):\penalty0 827--838,
  2010.

\bibitem[McLendon et~al.(2008)McLendon, Friedman, Bigner, Van~Meir, Brat,
  Mastrogianakis, Olson, Mikkelsen, Lehman, Aldape,
  et~al.]{mclendon2008comprehensive}
R.~McLendon, A.~Friedman, D.~Bigner, E.~G. Van~Meir, D.~J. Brat, G.~M.
  Mastrogianakis, J.~J. Olson, T.~Mikkelsen, N.~Lehman, K.~Aldape, et~al.
\newblock Comprehensive genomic characterization defines human glioblastoma
  genes and core pathways.
\newblock \emph{Nature}, 455\penalty0 (7216):\penalty0 1061--1068, 2008.

\bibitem[Metzker(2010)]{metzker2010sequencing}
M.~L. Metzker.
\newblock Sequencing technologies—the next generation.
\newblock \emph{Nature Reviews Genetics}, 11\penalty0 (1):\penalty0 31--46,
  2010.

\bibitem[Miyata et~al.(2003)Miyata, Koga, Kanda, Nishikido, Hayashi, and
  Kanetake]{miyata2003expression}
Y.~Miyata, S.~Koga, S.~Kanda, M.~Nishikido, T.~Hayashi, and H.~Kanetake.
\newblock Expression of cyclooxygenase-2 in renal cell carcinoma correlation
  with tumor cell proliferation, apoptosis, angiogenesis, expression of matrix
  metalloproteinase-2, and survival.
\newblock \emph{Clinical Cancer Research}, 9\penalty0 (5):\penalty0 1741--1749,
  2003.

\bibitem[Morris et~al.(2003)Morris, Hesson, Wagner, Morgan, Astuti, Lees,
  Cooper, Lee, Gentle, Macdonald, et~al.]{morris2003multigene}
M.~R. Morris, L.~B. Hesson, K.~J. Wagner, N.~V. Morgan, D.~Astuti, R.~D. Lees,
  W.~N. Cooper, J.~Lee, D.~Gentle, F.~Macdonald, et~al.
\newblock Multigene methylation analysis of wilms' tumour and adult renal cell
  carcinoma.
\newblock \emph{Oncogene}, 22\penalty0 (43):\penalty0 6794--6801, 2003.

\bibitem[Mortazavi et~al.(2008)Mortazavi, Williams, McCue, Schaeffer, and
  Wold]{mortazavi2008mapping}
A.~Mortazavi, B.~A. Williams, K.~McCue, L.~Schaeffer, and B.~Wold.
\newblock Mapping and quantifying mammalian transcriptomes by {RNA-Seq}.
\newblock \emph{Nature methods}, 5\penalty0 (7):\penalty0 621--628, 2008.

\bibitem[Oshlack et~al.(2010)Oshlack, Robinson, and Young]{oshlack2010rna}
A.~Oshlack, M.~D. Robinson, and M.~D. Young.
\newblock From {RNA-seq} reads to differential expression results.
\newblock \emph{Genome Biology}, 11\penalty0 (12):\penalty0 1--10, 2010.

\bibitem[Rapaport et~al.(2013)Rapaport, Khanin, Liang, Pirun, Krek, Zumbo,
  Mason, Socci, and Betel]{comprehensive}
F.~Rapaport, R.~Khanin, Y.~Liang, M.~Pirun, A.~Krek, P.~Zumbo, C.~E. Mason,
  N.~D. Socci, and D.~Betel.
\newblock Comprehensive evaluation of differential gene expression analysis
  methods for {RNA-seq} data.
\newblock \emph{Genome Biology}, 14\penalty0 (9):\penalty0 R95, 2013.

\bibitem[Risso et~al.(2014{\natexlab{a}})Risso, Ngai, Speed, and
  Dudoit]{risso2014normalization}
D.~Risso, J.~Ngai, T.~P. Speed, and S.~Dudoit.
\newblock Normalization of rna-seq data using factor analysis of control genes
  or samples.
\newblock \emph{Nature Biotechnology}, 32\penalty0 (9):\penalty0 896--902,
  2014{\natexlab{a}}.

\bibitem[Risso et~al.(2014{\natexlab{b}})Risso, Ngai, Speed, and
  Dudoit]{risso2014role}
D.~Risso, J.~Ngai, T.~P. Speed, and S.~Dudoit.
\newblock The role of spike-in standards in the normalization of {RNA-seq}.
\newblock In \emph{Statistical Analysis of Next Generation Sequencing Data},
  pages 169--190. Springer, 2014{\natexlab{b}}.

\bibitem[Roberts et~al.(2011)Roberts, Trapnell, Donaghey, Rinn, and
  Pachter]{roberts2011improving}
A.~Roberts, C.~Trapnell, J.~Donaghey, J.~L. Rinn, and L.~Pachter.
\newblock Improving {RNA-Seq} expression estimates by correcting for fragment
  bias.
\newblock \emph{Genome biology}, 12\penalty0 (3):\penalty0 1, 2011.

\bibitem[Robinson and Oshlack(2010)]{robinson2010scaling}
M.~D. Robinson and A.~Oshlack.
\newblock A scaling normalization method for differential expression analysis
  of {RNA-seq} data.
\newblock \emph{Genome Biology}, 11\penalty0 (3):\penalty0 1--9, 2010.

\bibitem[Robinson and Smyth(2007)]{robinson2007}
M.~D. Robinson and G.~K. Smyth.
\newblock Moderated statistical tests for assessing differences in tag
  abundance.
\newblock \emph{Bioinformatics}, 23\penalty0 (21):\penalty0 2881--2887, 2007.

\bibitem[Robinson et~al.(2010)Robinson, McCarthy, and Smyth]{edger}
M.~D. Robinson, D.~J. McCarthy, and G.~K. Smyth.
\newblock edger: a bioconductor package for differential expression analysis of
  digital gene expression data.
\newblock \emph{Bioinformatics}, 26\penalty0 (1):\penalty0 139--140, 2010.

\bibitem[Schena et~al.(1995)Schena, Shalon, Davis, and
  Brown]{schena1995quantitative}
M.~Schena, D.~Shalon, R.~W. Davis, and P.~O. Brown.
\newblock Quantitative monitoring of gene expression patterns with a
  complementary {DNA} microarray.
\newblock \emph{Science}, 270\penalty0 (5235):\penalty0 467, 1995.

\bibitem[Schurch et~al.(2016)Schurch, Schofield, Gierli{\'n}ski, Cole,
  Sherstnev, Singh, Wrobel, Gharbi, Simpson, Owen-Hughes,
  et~al.]{schurch2016many}
N.~J. Schurch, P.~Schofield, M.~Gierli{\'n}ski, C.~Cole, A.~Sherstnev,
  V.~Singh, N.~Wrobel, K.~Gharbi, G.~G. Simpson, T.~Owen-Hughes, et~al.
\newblock How many biological replicates are needed in an {RNA-seq} experiment
  and which differential expression tool should you use?
\newblock \emph{RNA}, 22\penalty0 (6):\penalty0 839--851, 2016.

\bibitem[{SEQC/MAQC-III Consortium}(2014)]{seqc}
{SEQC/MAQC-III Consortium}.
\newblock A comprehensive assessment of {RNA-seq} accuracy, reproducibility and
  information content by the {Sequencing Quality Control Consortium}.
\newblock \emph{Nature Biotechnology}, 32\penalty0 (9):\penalty0 903--914,
  2014.

\bibitem[Shi et~al.(2006)Shi, Reid, Jones, Shippy, Warrington, Baker, Collins,
  De~Longueville, Kawasaki, Lee, et~al.]{maqc}
L.~Shi, L.~H. Reid, W.~D. Jones, R.~Shippy, J.~A. Warrington, S.~C. Baker,
  P.~J. Collins, F.~De~Longueville, E.~S. Kawasaki, K.~Y. Lee, et~al.
\newblock The microarray quality control (maqc) project shows inter-and
  intraplatform reproducibility of gene expression measurements.
\newblock \emph{Nature Biotechnology}, 24\penalty0 (9):\penalty0 1151--1161,
  2006.

\bibitem[Smyth and Verbyla(1996)]{smyth1996conditional}
G.~Smyth and A.~Verbyla.
\newblock A conditional likelihood approach to residual maximum likelihood
  estimation in generalized linear models.
\newblock \emph{J. R. Stat. Soc: Series B}, 58\penalty0 (3):\penalty0 565--572,
  1996.

\bibitem[Soneson and Delorenzi(2013)]{soneson2013comparison}
C.~Soneson and M.~Delorenzi.
\newblock A comparison of methods for differential expression analysis of
  {RNA-seq} data.
\newblock \emph{BMC Bioinformatics}, 14:\penalty0 91, 2013.

\bibitem[Tang et~al.(2009)Tang, Chang, Su, Chen, Lai, Wu, Hsu, Lin, Lai, and
  Lin]{tang2009myc}
S.-W. Tang, W.-H. Chang, Y.-C. Su, Y.-C. Chen, Y.-H. Lai, P.-T. Wu, C.-I. Hsu,
  W.-C. Lin, M.-K. Lai, and J.-Y. Lin.
\newblock Myc pathway is activated in clear cell renal cell carcinoma and
  essential for proliferation of clear cell renal cell carcinoma cells.
\newblock \emph{Cancer Letters}, 273\penalty0 (1):\penalty0 35--43, 2009.

\bibitem[Trudu et~al.(2013)Trudu, Janas, Lanzani, Debaix, Schaeffer, Ikehata,
  Citterio, Demaretz, Trevisani, Ristagno, et~al.]{trudu2013common}
M.~Trudu, S.~Janas, C.~Lanzani, H.~Debaix, C.~Schaeffer, M.~Ikehata,
  L.~Citterio, S.~Demaretz, F.~Trevisani, G.~Ristagno, et~al.
\newblock Common noncoding umod gene variants induce salt-sensitive
  hypertension and kidney damage by increasing uromodulin expression.
\newblock \emph{Nature Medicine}, 19\penalty0 (12):\penalty0 1655--1660, 2013.

\bibitem[Tun et~al.(2010)Tun, Marlow, Von~Roemeling, Cooper, Kreinest, Wu,
  Luxon, Sinha, Anastasiadis, and Copland]{tun2010pathway}
H.~W. Tun, L.~A. Marlow, C.~A. Von~Roemeling, S.~J. Cooper, P.~Kreinest, K.~Wu,
  B.~A. Luxon, M.~Sinha, P.~Z. Anastasiadis, and J.~A. Copland.
\newblock Pathway signature and cellular differentiation in clear cell renal
  cell carcinoma.
\newblock \emph{PloS One}, 5\penalty0 (5):\penalty0 e10696, 2010.

\bibitem[Wang et~al.(2010)Wang, Feng, Wang, Wang, and Zhang]{wang2010degseq}
L.~Wang, Z.~Feng, X.~Wang, X.~Wang, and X.~Zhang.
\newblock {DEGseq}: an {R} package for identifying differentially expressed
  genes from {RNA-seq} data.
\newblock \emph{Bioinformatics}, 26\penalty0 (1):\penalty0 136--138, 2010.

\bibitem[Wang et~al.(2009)Wang, Gerstein, and Snyder]{wang2009rna}
Z.~Wang, M.~Gerstein, and M.~Snyder.
\newblock {RNA-Seq}: a revolutionary tool for transcriptomics.
\newblock \emph{Nature Reviews Genetics}, 10\penalty0 (1):\penalty0 57--63,
  2009.

\bibitem[West(2003)]{West03bayesianfactor}
M.~West.
\newblock {B}ayesian factor regression models in the ``large $p$, small $n$''
  paradigm.
\newblock In \emph{Bayesian Statistics}, 2003.

\bibitem[Xu et~al.(2013)Xu, Su, Hong, Thierry-Mieg, Thierry-Mieg, Kreil, Mason,
  Tong, and Shi]{xu2013cross}
J.~Xu, Z.~Su, H.~Hong, J.~Thierry-Mieg, D.~Thierry-Mieg, D.~P. Kreil, C.~E.
  Mason, W.~Tong, and L.~Shi.
\newblock Cross-platform ultradeep transcriptomic profiling of human reference
  {RNA} samples by {RNA-Seq}.
\newblock \emph{Scientific Data}, 1:\penalty0 140020--140020, 2013.

\bibitem[Yamasaki et~al.(2012)Yamasaki, Seki, Yamada, Yoshino, Hidaka,
  Chiyomaru, Nohata, Kinoshita, Nakagawa, and Enokida]{yamasaki2012tumor}
T.~Yamasaki, N.~Seki, Y.~Yamada, H.~Yoshino, H.~Hidaka, T.~Chiyomaru,
  N.~Nohata, T.~Kinoshita, M.~Nakagawa, and H.~Enokida.
\newblock Tumor suppressive microrna-138 contributes to cell migration and
  invasion through its targeting of vimentin in renal cell carcinoma.
\newblock \emph{International Journal of Oncology}, 41\penalty0 (3):\penalty0
  805--817, 2012.

\bibitem[Zhang et~al.(2014)Zhang, Jhaveri, Marshall, Bauer, Edson, Narayanan,
  Robinson, Lundberg, Bartlett, Wray, et~al.]{zhang2014comparative}
Z.~H. Zhang, D.~J. Jhaveri, V.~M. Marshall, D.~C. Bauer, J.~Edson, R.~K.
  Narayanan, G.~J. Robinson, A.~E. Lundberg, P.~F. Bartlett, N.~R. Wray, et~al.
\newblock A comparative study of techniques for differential expression
  analysis on {RNA-Seq} data.
\newblock \emph{PloS one}, 9\penalty0 (8):\penalty0 e103207, 2014.

\bibitem[Zhou et~al.(2010)Zhou, Chen, Li, Li, Hu, Huang, Zhao, Liang, Wang,
  Sun, et~al.]{zhou2010integrated}
L.~Zhou, J.~Chen, Z.~Li, X.~Li, X.~Hu, Y.~Huang, X.~Zhao, C.~Liang, Y.~Wang,
  L.~Sun, et~al.
\newblock Integrated profiling of micrornas and mrnas: micrornas located on
  xq27. 3 associate with clear cell renal cell carcinoma.
\newblock \emph{PloS One}, 5\penalty0 (12):\penalty0 e15224, 2010.

\bibitem[Zhou and Carin(2015)]{NBP2012}
M.~Zhou and L.~Carin.
\newblock Negative binomial process count and mixture modeling.
\newblock \emph{IEEE Trans. Pattern Anal. Mach. Intell.}, 37\penalty0
  (2):\penalty0 307--320, 2015.

\bibitem[Zhou et~al.(2012)Zhou, Hannah, Dunson, and
  Carin]{BNBP_PFA_AISTATS2012}
M.~Zhou, L.~Hannah, D.~Dunson, and L.~Carin.
\newblock Beta-negative binomial process and {P}oisson factor analysis.
\newblock In \emph{AISTATS}, pages 1462--1471, 2012.

\bibitem[Zhou et~al.(2016)Zhou, Padilla, and Scott]{NBP_CountMatrix}
M.~Zhou, O.~H.~M. Padilla, and J.~G. Scott.
\newblock Priors for random count matrices derived from a family of negative
  binomial processes.
\newblock \emph{J. Amer. Statist. Assoc.}, 111\penalty0 (515):\penalty0
  1144--1156, 2016.

\bibitem[Zyprych-Walczak et~al.(2015)Zyprych-Walczak, Szabelska, Handschuh,
  G{\'o}rczak, Klamecka, Figlerowicz, and Siatkowski]{zyprych}
J.~Zyprych-Walczak, A.~Szabelska, L.~Handschuh, K.~G{\'o}rczak, K.~Klamecka,
  M.~Figlerowicz, and I.~Siatkowski.
\newblock The impact of normalization methods on {RNA-Seq} data analysis.
\newblock \emph{BioMed Research International}, 2015, 2015.

\end{thebibliography}
\end{spacing}

\bigskip
\normalsize
\appendix
\newpage

\begin{center}
\Large{Supplementary Material for BNP-Seq: Bayesian Nonparametric Differential Expression Analysis of Sequencing Count Data}
\end{center}

\section{Chinese restaurant table (CRT) distribution}

The negative binomial distribution $m \sim \mbox{NB}(r,p)$ with the probability mass function $$f_M(m)=\frac{\Gamma(m+r)}{m! \Gamma(r)}(1-p)^r p^m,~~m\in\{0,1,\ldots\}$$ can be augmented as a gamma mixed Poisson distribution as $$m \sim \mbox{Pois}(\lambda), ~\lambda \sim \mbox{Gamma}(r,p/(1-p)),$$ where the gamma distribution is parametrized by its shape $r$ and scale $p/(1-p)$. It can be augmented under a compound Poisson representation as $$m=\sum_{t=1}^{\ell} u_t, ~u_t \sim \mbox{Log}(p), ~\ell \sim \mbox{Pois}(-r \ln(1-p)),$$ where $u \sim \mbox{Log}(p)$ is the logarithmic distribution with probability generation function  $C_U(z)=\ln (1-pz)/ \ln(1-p)$, $|z|<p^{-1}$. As in \citet{NBP2012}, we denote the conditional posterior distribution of $\ell$ given $m$ and $r$ by $(\ell\given m, r) \sim \mbox{CRT}(m,r)$ and sample it with the summation of independent Bernoulli random variables as $\ell=\sum_{n=1}^m b_n$, $b_n \sim \mbox{Bernoulli}[r/(n-1+r)]$.

\section{Additional tables and figures}

\begin{table}[h]
	\centering
	\caption{AUC-ROC in the GNBP simulation setup for different true fold changes.}
	\small
	\begin{tabular}{ l||c c c c }
		&\multicolumn{4}{c}{Fold change} \\
		\hline \hline
		Method & 1.4 & 1.6 & 1.8 & 2\\
		\hline
		GNBP & \textbf{0.9226} $\pm$ 0.006 & \textbf{0.9625} $\pm$ 0.003 & 0.9777 $\pm$ 0.003 & 0.9864 $\pm$ 0.002 \\
		BNBP & 0.9156 $\pm$ 0.005 & 0.9610 $\pm$ 0.003 & \textbf{0.9783} $\pm$ 0.002 & \textbf{0.9875} $\pm$ 0.002 \\
		edgeR & 0.9004 $\pm$ 0.007 & 0.9463 $\pm$ 0.004 & 0.9653 $\pm$ 0.003 & 0.9778 $\pm$ 0.003 \\
		DESeq & 0.8986 $\pm$ 0.008 & 0.9444 $\pm$ 0.004 & 0.9634 $\pm$ 0.003 & 0.9764 $\pm$ 0.003 \\
		baySeq & 0.7542 $\pm$ 0.008 & 0.8247 $\pm$ 0.012 & 0.8752 $\pm$ 0.003 & 0.9114 $\pm$ 0.008 \\
		NBP & 0.9035 $\pm$ 0.007 & 0.9476 $\pm$ 0.004 & 0.9665 $\pm$ 0.003 & 0.9786 $\pm$ 0.003 \\
		NBPscaled & 0.8596 $\pm$ 0.014 & 0.8990 $\pm$ 0.017 & 0.9366 $\pm$ 0.009 & 0.9506 $\pm$ 0.0053 \\
		\hline
	\end{tabular}	
\end{table}

\begin{table}[h]
	\centering
	\caption{AUC-PR in the GNBP simulation setup for different true fold changes.}
	\small
	\begin{tabular}{ l||c c c c }
		&\multicolumn{4}{c}{Fold change} \\
		\hline \hline
		Method & 1.4 & 1.6 & 1.8 & 2\\
		\hline
		GNBP & 0.7873 $\pm$ 0.011 & \textbf{0.8998} $\pm$ 0.006 & \textbf{0.9382} $\pm$ 0.003 & \textbf{0.9607} $\pm$ 0.003 \\
		BNBP & 0.5660 $\pm$ 0.011 & 0.8189 $\pm$ 0.008 & 0.9213 $\pm$ 0.005 & 0.9563 $\pm$ 0.002 \\
		edgeR & 0.7857 $\pm$ 0.015 & 0.8742 $\pm$ 0.007 & 0.9136 $\pm$ 0.003 & 0.9403 $\pm$ 0.003 \\
		DESeq & 0.7848 $\pm$ 0.014 & 0.8714 $\pm$ 0.007 & 0.9107 $\pm$ 0.002 & 0.9369 $\pm$ 0.004 \\
		baySeq & 0.6517 $\pm$ 0.012 & 0.7655 $\pm$ 0.015 & 0.8329 $\pm$ 0.003 & 0.8756 $\pm$ 0.004 \\
		NBP & \textbf{0.7934} $\pm$ 0.014 & 0.8770 $\pm$ 0.007 & 0.9156 $\pm$ 0.003 & 0.9399 $\pm$ 0.003 \\
		NBPscaled & 0.6822 $\pm$ 0.035 & 0.7515 $\pm$ 0.036 & 0.8298 $\pm$ 0.028 & 0.8533 $\pm$ 0.012 \\ 
		\hline
	\end{tabular}	
\end{table}

\begin{table}[h]
	\centering
	\caption{AUC-ROC in the BNBP simulation setup for different true fold changes.}
	\small
	\begin{tabular}{ l||c c c c }
		&\multicolumn{4}{c}{Fold change} \\
		\hline \hline
		Method & 1.4 & 1.6 & 1.8 & 2\\
		\hline
		GNBP & \textbf{0.9648} $\pm$ 0.001 & \textbf{0.9847} $\pm$ 0.001 & \textbf{0.9914} $\pm$ 0.0014 & \textbf{0.9968} $\pm$ 0.001 \\
		BNBP & 0.9635 $\pm$ 0.001 & \textbf{0.9848} $\pm$ 0.002 & \textbf{0.9922} $\pm$ 0.0009 & \textbf{0.9971} $\pm$ 0.0009 \\
		edgeR &0.9399 $\pm$ 0.001 & 0.9706 $\pm$ 0.003 & 0.9829 $\pm$ 0.0017 & 0.9929 $\pm$ 0.00189 \\
		DESeq & 0.9383 $\pm$ 0.002 & 0.9694 $\pm$ 0.003 & 0.9818 $\pm$ 0.0016 & 0.9920 $\pm$ 0.0018 \\
		baySeq & 0.7919 $\pm$ 0.007 & 0.8699 $\pm$ 0.07 & 0.9167 $\pm$ 0.007 & 0.9590 $\pm$ 0.0041 \\
		NBP & 0.9438 $\pm$ 0.001 & 0.9729 $\pm$ 0.003 & 0.9844 $\pm$ 0.002 & 0.9935 $\pm$ 0.0016 \\
		NBPscaled & 0.8939 $\pm$ 0.0107 & 0.9499 $\pm$ 0.0092 & 0.9606 $\pm$ 0.0094 & 0.9811 $\pm$ 0.008 \\ 
		\hline
	\end{tabular}	
\end{table}

\begin{table}[h]
	\centering
	\caption{AUC-PR in the BNBP simulation setup for different true fold changes.}
	\small
	\begin{tabular}{ l||c c c c }
		&\multicolumn{4}{c}{Fold change} \\
		\hline \hline
		Method & 1.4 & 1.6 & 1.8 & 2\\
		\hline
		GNBP & 0.8632 $\pm$ 0.011 & \textbf{0.9431} $\pm$ 0.005 & \textbf{0.9703} $\pm$ 0.003 & \textbf{0.9881} $\pm$ 0.002 \\
		BNBP & 0.8356 $\pm$ 0.012 & \textbf{0.9432} $\pm$ 0.003 & \textbf{0.9725} $\pm$ 0.002 & \textbf{0.9889} $\pm$ 0.003 \\
		edgeR & 0.8674 $\pm$ 0.006 & 0.9275 $\pm$ 0.005 & 0.9557 $\pm$ 0.003 & 0.9783 $\pm$ 0.004 \\
		DESeq & 0.8634 $\pm$ 0.004 & 0.9240 $\pm$ 0.005 & 0.9523 $\pm$ 0.003 & 0.9759 $\pm$ 0.003 \\ 
		baySeq & 0.7413 $\pm$ 0.015 & 0.8408 $\pm$ 0.01 & 0.8963 $\pm$ 0.007 & 0.9434 $\pm$ 0.003 \\
		NBP & \textbf{0.8708} $\pm$ 0.006 & 0.9302 $\pm$ 0.005 & 0.9577 $\pm$ 0.003 & 0.9798 $\pm$ 0.003 \\
		NBPscaled & 0.7450 $\pm$ 0.03 & 0.8648 $\pm$ 0.019 & 0.8846 $\pm$ 0.028 & 0.9318 $\pm$ 0.025 \\ 
		\hline
	\end{tabular}	
\end{table}

\begin{table}[h]
	\centering
	\caption{AUC-ROC in the baySeq simulation setup for different true fold changes.}
	\small
	\begin{tabular}{ l||c c c c }
		&\multicolumn{4}{c}{Fold change} \\
		\hline \hline
		Method & 1.4 & 1.6 & 1.8 & 2\\
		\hline
		GNBP & 0.8772 $\pm$ 0.009 & 0.9286 $\pm$ 0.005 & 0.9585 $\pm$ 0.004 & 0.9738 $\pm$ 0.001 \\
		BNBP & \textbf{0.8823} $\pm$ 0.005 & \textbf{0.9382} $\pm$ 0.004 & \textbf{0.9674} $\pm$ 0.003 & \textbf{0.9812} $\pm$ 0.0015 \\
		edgeR & 0.8702 $\pm$ 0.008 & 0.9216 $\pm$ 0.0042 & 0.9518 $\pm$ 0.004 & 0.9687 $\pm$ 0.003 \\
		DESeq & 0.8705 $\pm$ 0.0083 & 0.9220 $\pm$ 0.004 & 0.9520 $\pm$ 0.0036 & 0.9688 $\pm$ 0.003 \\
		baySeq & 0.7222 $\pm$ 0.0089 & 0.7887 $\pm$ 0.0067 & 0.8489 $\pm$ 0.012 & 0.8911 $\pm$ 0.0099 \\
		NBP & 0.8769 $\pm$ 0.0075 & 0.9270 $\pm$ 0.0045 & 0.9567 $\pm$ 0.0031 & 0.9725 $\pm$ 0.0026 \\
		NBPscaled & 0.8752 $\pm$ 0.009 & 0.9248 $\pm$ 0.0044 & 0.9571 $\pm$ 0.0071 & 0.9719 $\pm$ 0.0031 \\ 
		\hline
	\end{tabular}	
\end{table}

\begin{table}[h]
	\centering
	\caption{AUC-PR in the baySeq simulation setup for different true fold changes.}
	\small
	\begin{tabular}{ l||c c c c}
		&\multicolumn{4}{c}{Fold change} \\
		\hline \hline
		Method & 1.4 & 1.6 & 1.8 & 2\\
		\hline
		GNBP & \textbf{0.7194} $\pm$ 0.015 & \textbf{0.8372} 0.0095 $\pm$ & \textbf{0.8984} $\pm$ 0.0074 & \textbf{0.9332} $\pm$ 0.0041 \\
		BNBP & 0.5733 $\pm$ 0.012 & 0.7448 $\pm$ 0.014 & 0.8826 $\pm$ 0.0055 & \textbf{0.9337} $\pm$ 0.0058 \\
		edgeR & 0.7004 $\pm$ 0.013 & 0.8152 $\pm$ 0.008 & 0.8787 $\pm$ 0.0066 & 0.9173 $\pm$ 0.0054 \\
		DESeq & 0.7042 $\pm$ 0.013 & 0.8180 $\pm$ 0.0082 & 0.8813 $\pm$ 0.0058 & 0.9194 $\pm$ 0.005 \\
		baySeq & 0.5806 $\pm$ 0.0096 & 0.7034 $\pm$ 0.0057 & 0.7877 $\pm$ 0.0104 & 0.8482 $\pm$ 0.0106 \\
		NBP & \textbf{0.7223} $\pm$ 0.0129 & 0.8312 $\pm$ 0.0075 & 0.8913 $\pm$ 0.0058 & 0.9248 $\pm$ 0.0055 \\
		NBPscaled & 0.7203 $\pm$ 0.0155 & 0.8333 $\pm$ 0.006 & \textbf{0.8940} $\pm$ 0.012 & 0.9256 $\pm$ 0.0047 \\
		\hline
	\end{tabular}	
\end{table}

\begin{table}[h]
	\centering
	\caption{Area under the ROC curve for the range with $FPR \leq 0.1$ and area under the PR curve for the range with $\mbox{Recall} \leq 0.1$ for both the PSU and BGI datasets, with the log2 cut-off value fixed at 2.}
	\small
	\begin{tabular}{ l||c c | c c}
		&\multicolumn{2}{c}{PSU} & \multicolumn{2}{c}{BGI} \\
		\hline \hline
		Method & AUCroc & AUCpr & AUCroc & AUCpr\\
		\hline
		GNBP & \textbf{0.0627} &  \textbf{0.0980} & \textbf{0.0716} & \textbf{0.0995} \\
		BNBP & \textbf{0.0628} &  \textbf{0.0980} & 0.0685 &  0.0986 \\
		edgeR & 0.0587 & \textbf{0.0980} & 0.0527 &  \textbf{0.0995} \\
		DESeq & 0.0514 &  \textbf{0.0980} & 0.0521 & \textbf{0.0995} \\
		baySeq & 0.0533 & \textbf{0.0980} & 0.0258 & 0.0757 \\
		NBP  &  0.0356 &  0.0921 & 0.0390 &  0.0968 \\
		NBPscaled & 0.0404  & 0.0911 & 0.0369 & 0.0957\\
		\hline
	\end{tabular}
	\label{table:AUC0.1}	
\end{table}

\begin{figure}[h] 
	\centering
	\begin{subfigure}[b]{0.88\textwidth}
		\includegraphics[width=1\textwidth]{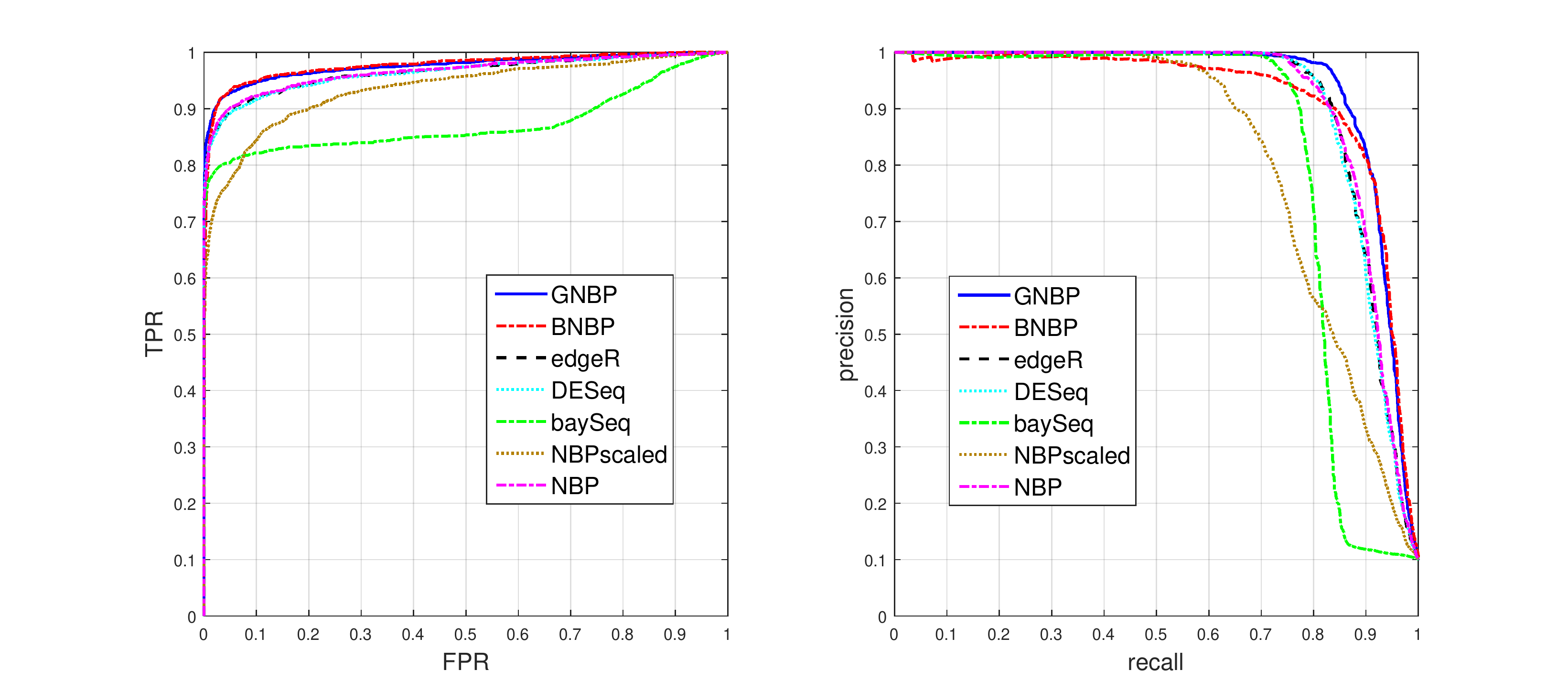}
		\caption{GNBP setup}
	\end{subfigure}
	
	\begin{subfigure}[b]{0.88\textwidth}
		\includegraphics[width=1\textwidth]{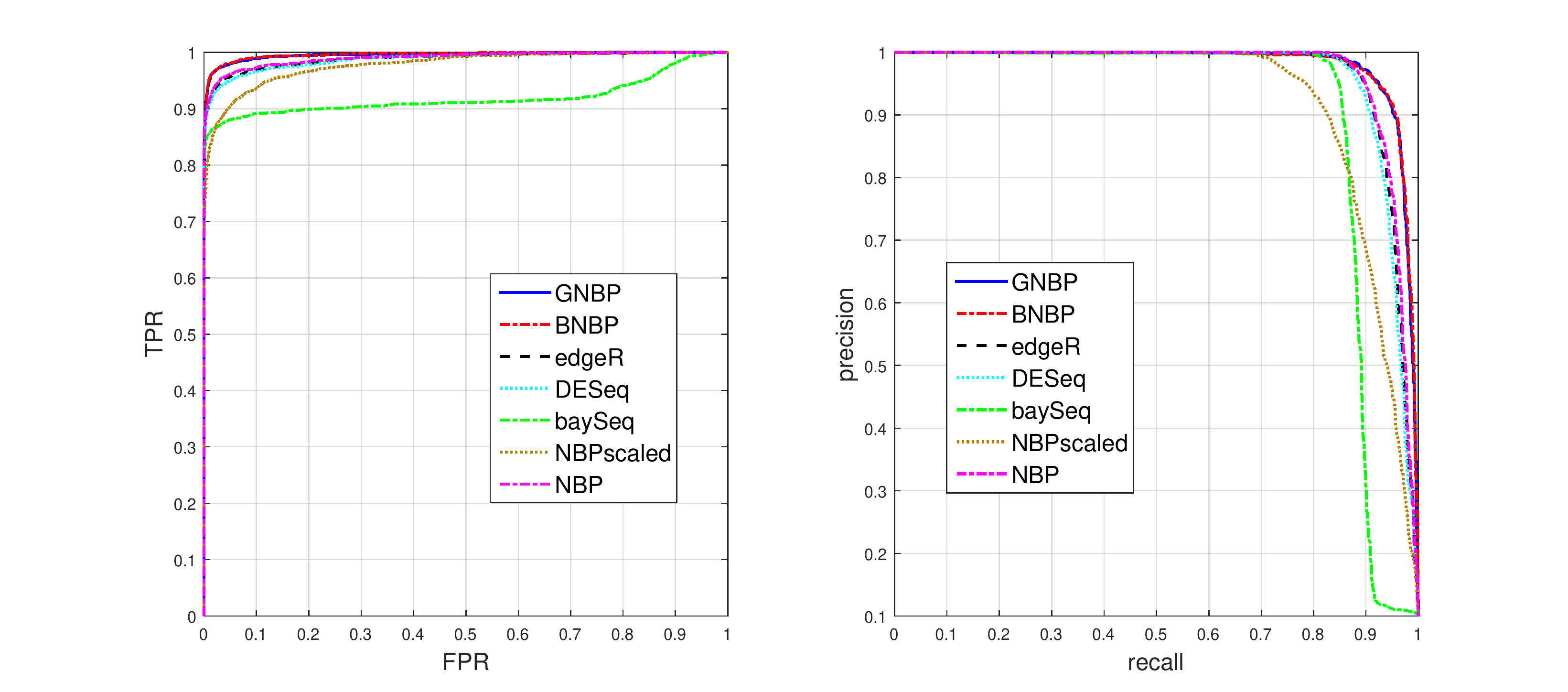}
		\caption{BNBP setup}
	\end{subfigure}
	
	\begin{subfigure}[b]{0.88\textwidth}
		\includegraphics[width=1\textwidth]{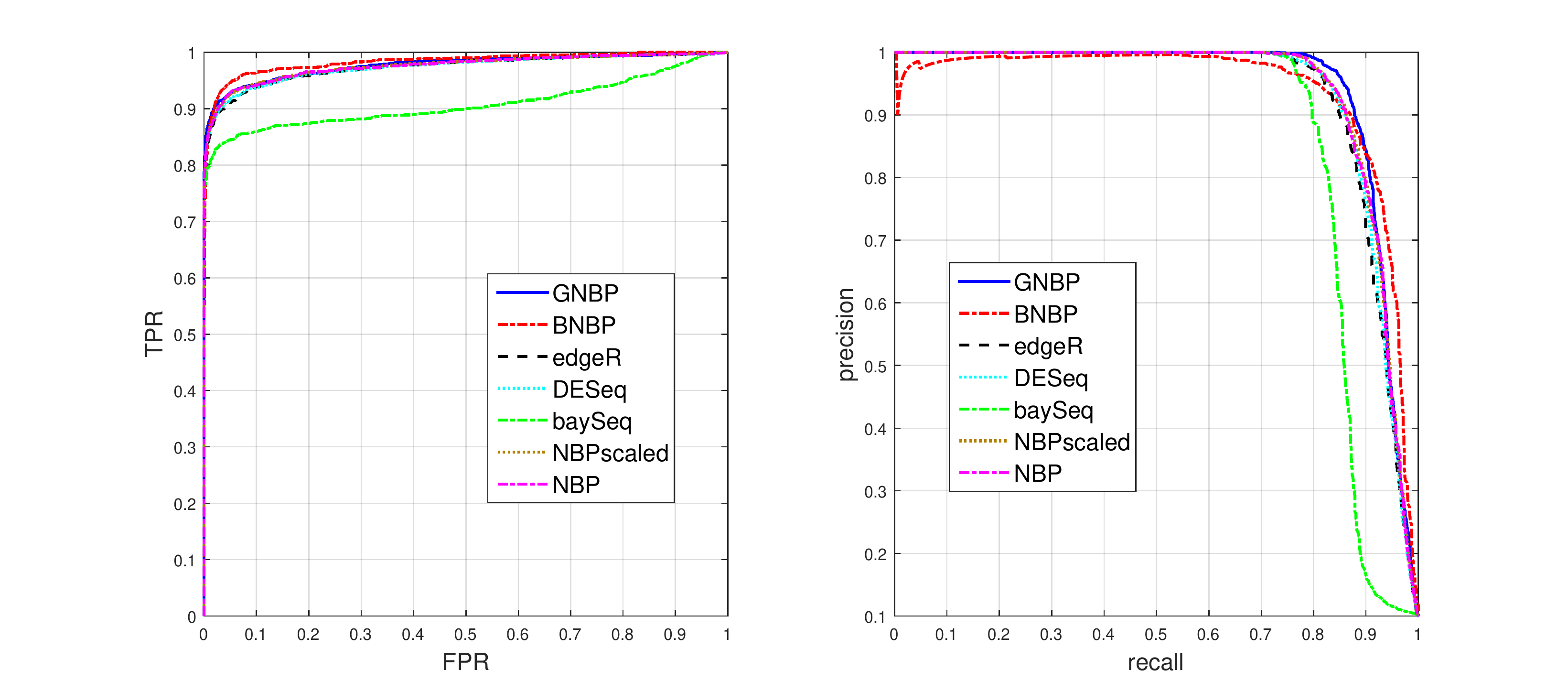}
		\caption{baySeq setup}
	\end{subfigure}
	\caption{\textbf{left column}: \small ROC curve, \textbf{right column}: PR curve. Performance of different methods in detecting the differential expression of simulated data generated from different setups with a fold change of 1.8 for truly differentially expressed genes.}
	\label{fig:simulated}
\end{figure}

\begin{figure}[h]
	\centering
	\includegraphics[width=1.\textwidth]{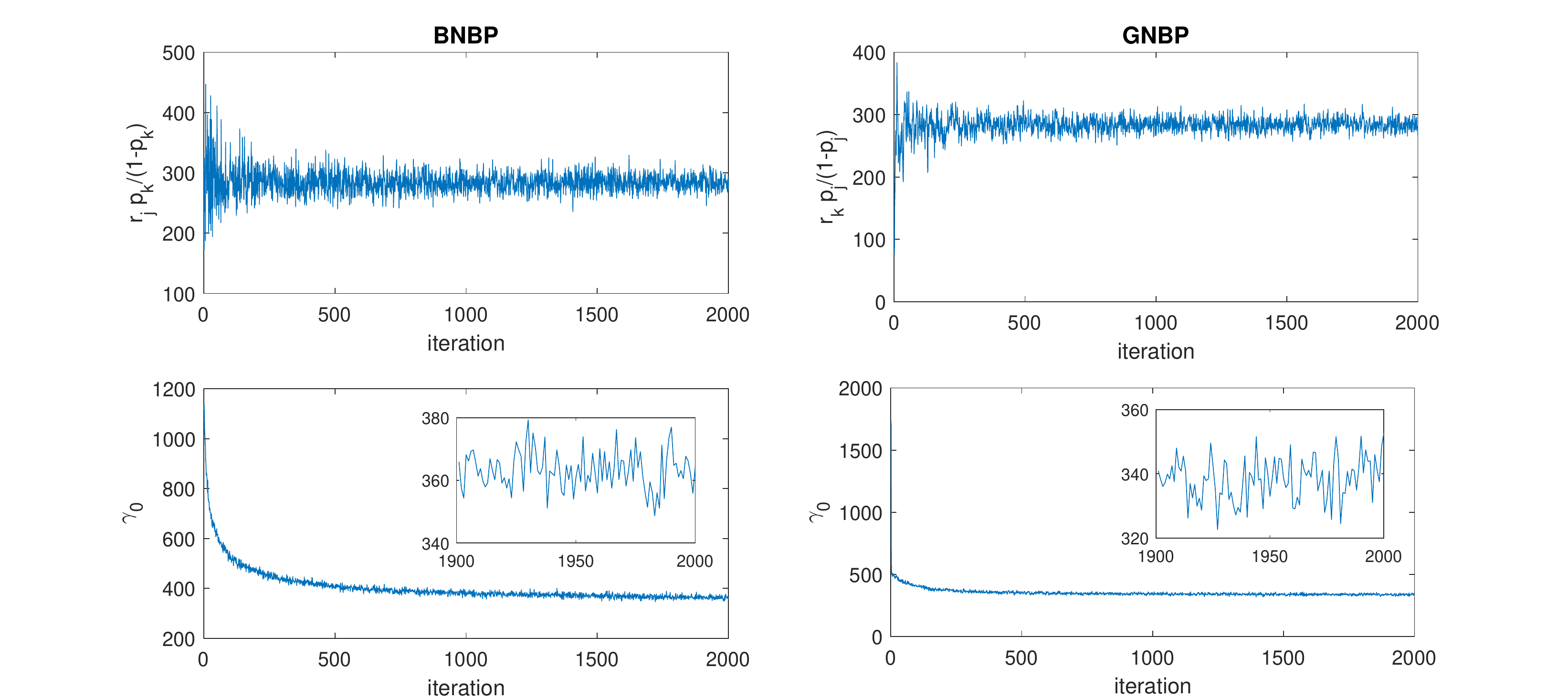}
	\caption{Trace plots of 2000 MCMC samples for example parameters of the BNBP (left column) and GNBP (right column) methods, applied to the BGI dataset.}
	\label{fig:trace-plot}
\end{figure}

\begin{figure}[!t]
	\centering
	\begin{subfigure}[b]{0.86\textwidth}
		\includegraphics[width=1.05\textwidth]{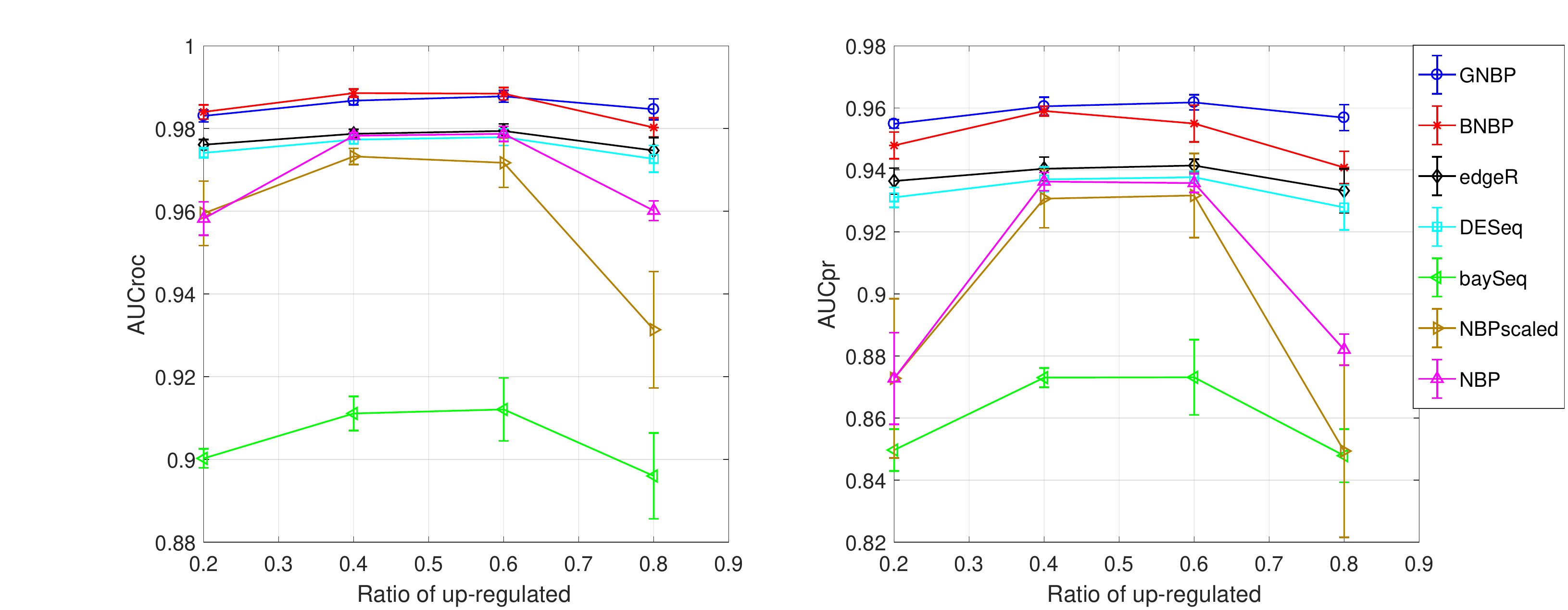}
		\caption{Varying up/down-regulation proportions}
	\end{subfigure}
	\begin{subfigure}[b]{0.86\textwidth}
		\includegraphics[width=1.05\textwidth]{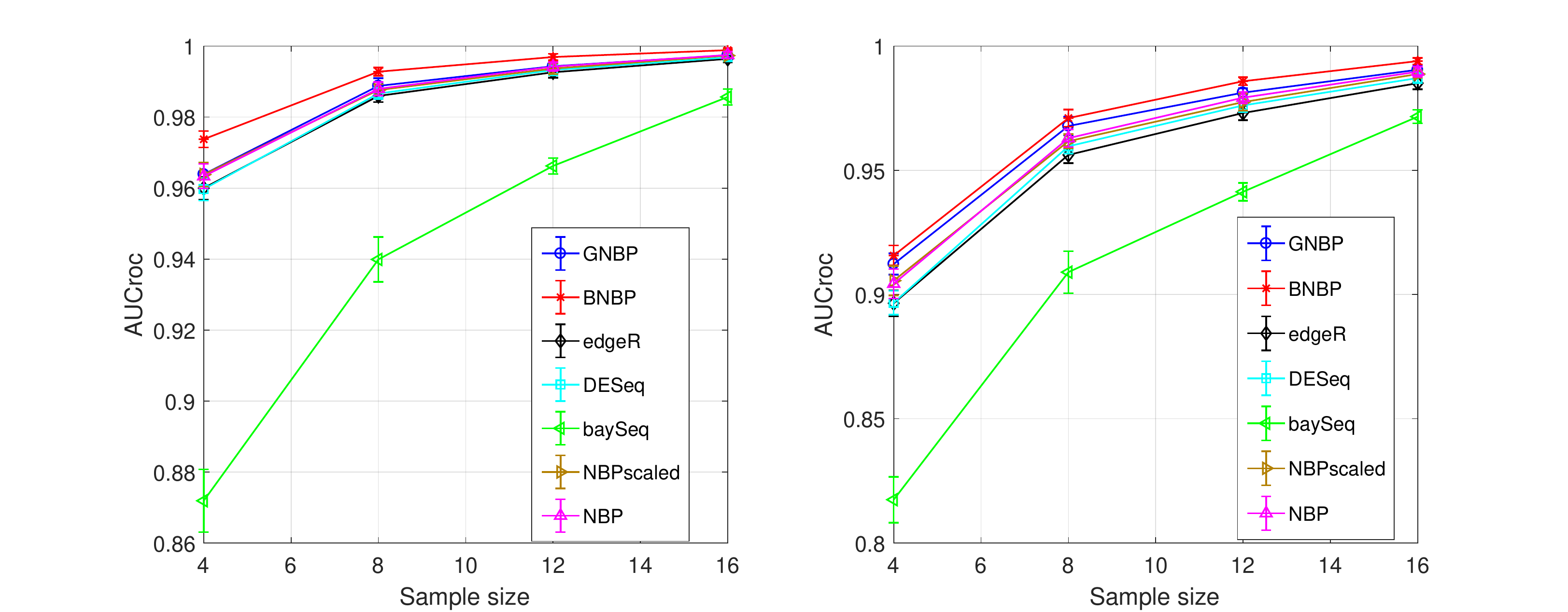}
		\caption{Different sample sizes}
	\end{subfigure}
	\begin{subfigure}[b]{0.86\textwidth}
		\includegraphics[width=1.05\textwidth]{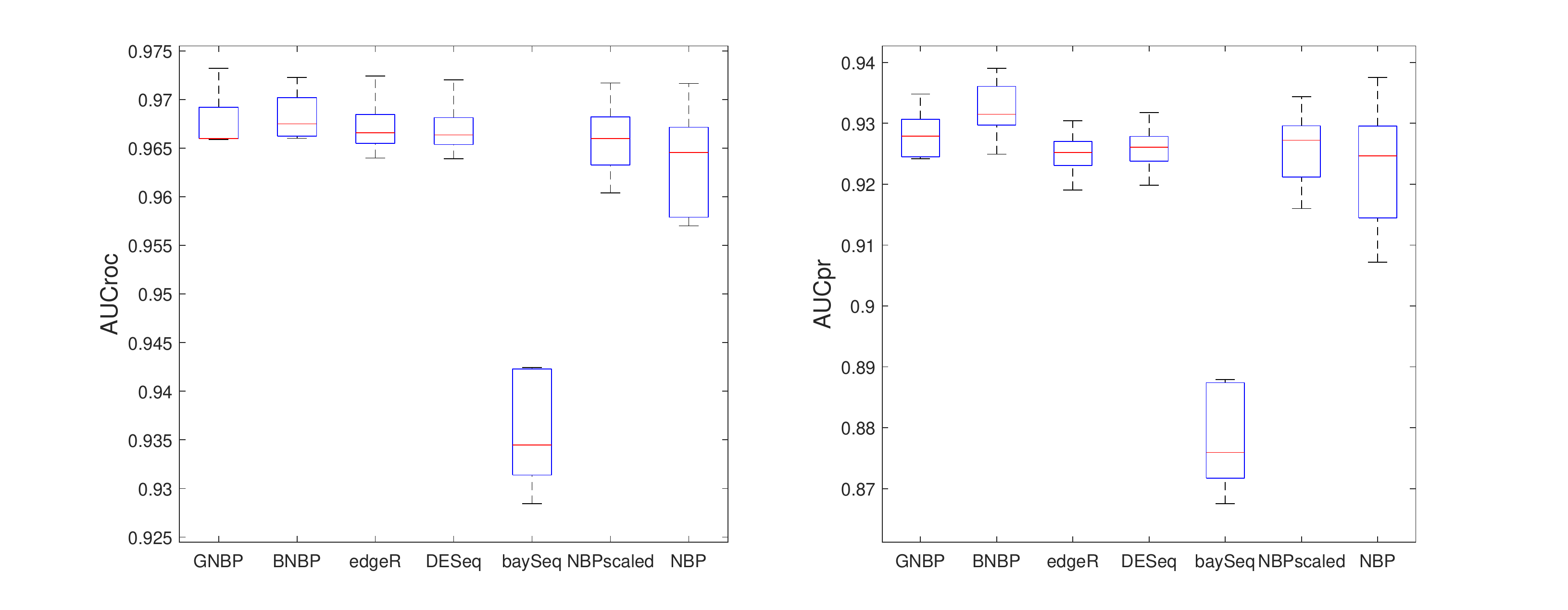}
		\caption{Continuous true fold changes}
	\end{subfigure}
	\caption{\textbf{left column}: \small AUC-ROC values, \textbf{right column}: AUC-PR values. Performance comparison of different methods in detecting differentially expressed genes under various scenarios using synthetic data generated with baySeq. (a) The proportion of up-regulated genes in true differentially expressed genes increases from 20\% to 80\% with 20\% increments. (b) The sample size in each group is increased from 4 to 16 with increments of size 4. (c) The true fold change of differentially expressed genes is sampled from a uniform distribution in the interval $[1.4,2]$.
	}
	\label{fig:synthetic2}
\end{figure}

\begin{figure}[h] 
	\centering
	\begin{subfigure}[b]{0.7\textwidth}
		\includegraphics[width=1\textwidth]{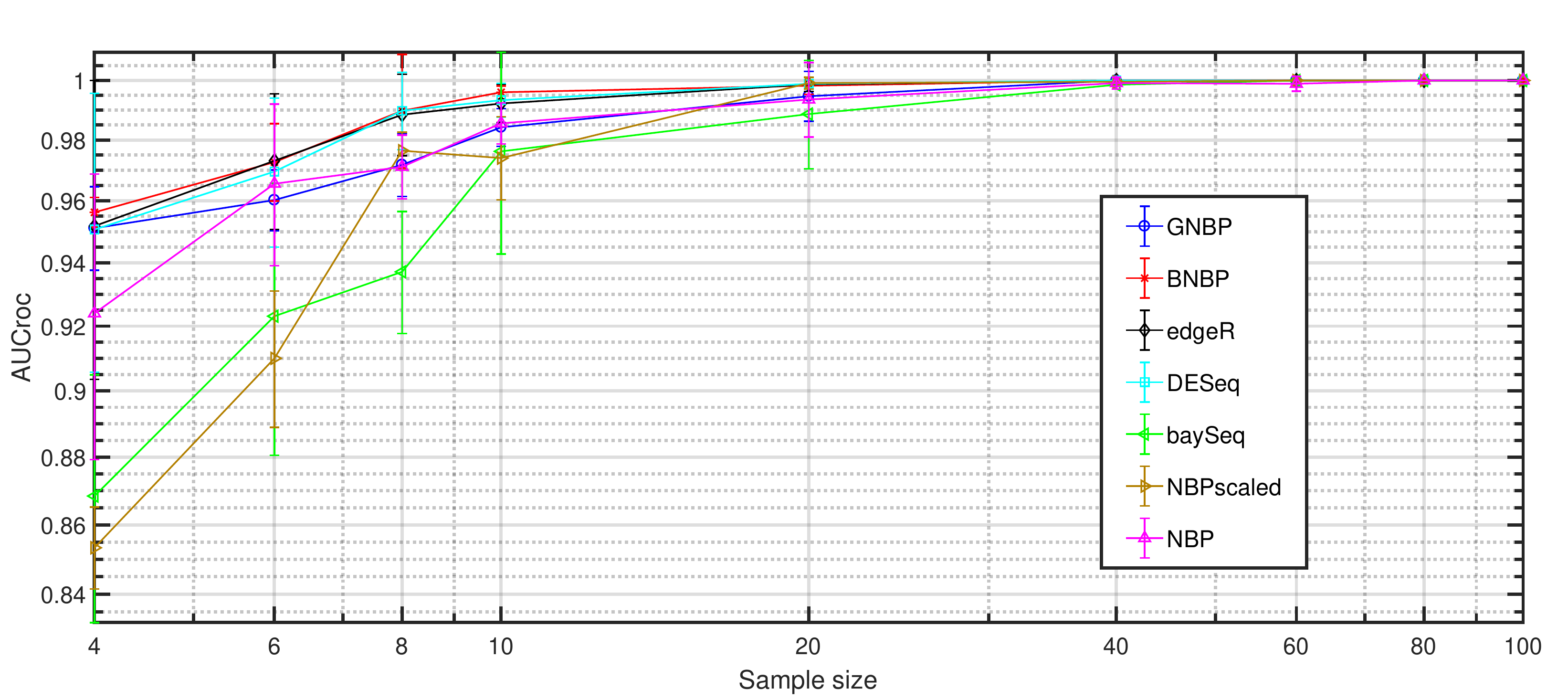}
		\caption{AUC-ROC}
	\end{subfigure}
	
	\begin{subfigure}[b]{0.7\textwidth}
		\includegraphics[width=1\textwidth]{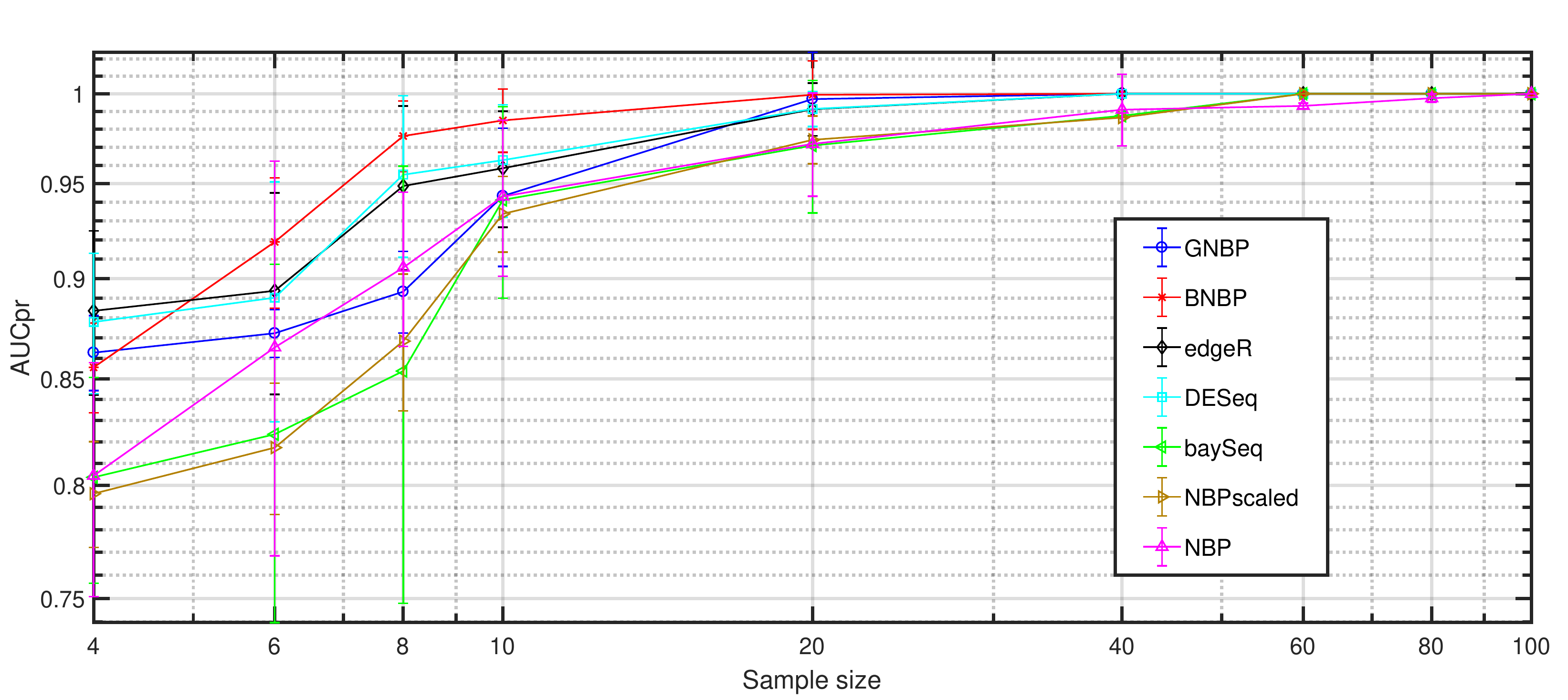}
		\caption{AUC-PR}
	\end{subfigure}
	
	\caption{\textbf{(a)} \small AUC-ROC and \textbf{(b)} AUC-PR in the baySeq simulation setup with 100 genes and different sample sizes, where 10 genes are equally likely to be up- or down-regulated with a fold change of 2.}
	\label{fig:small}
\end{figure}

\begin{figure}[!t] 
	\centering
	\includegraphics[width=0.6\textwidth]{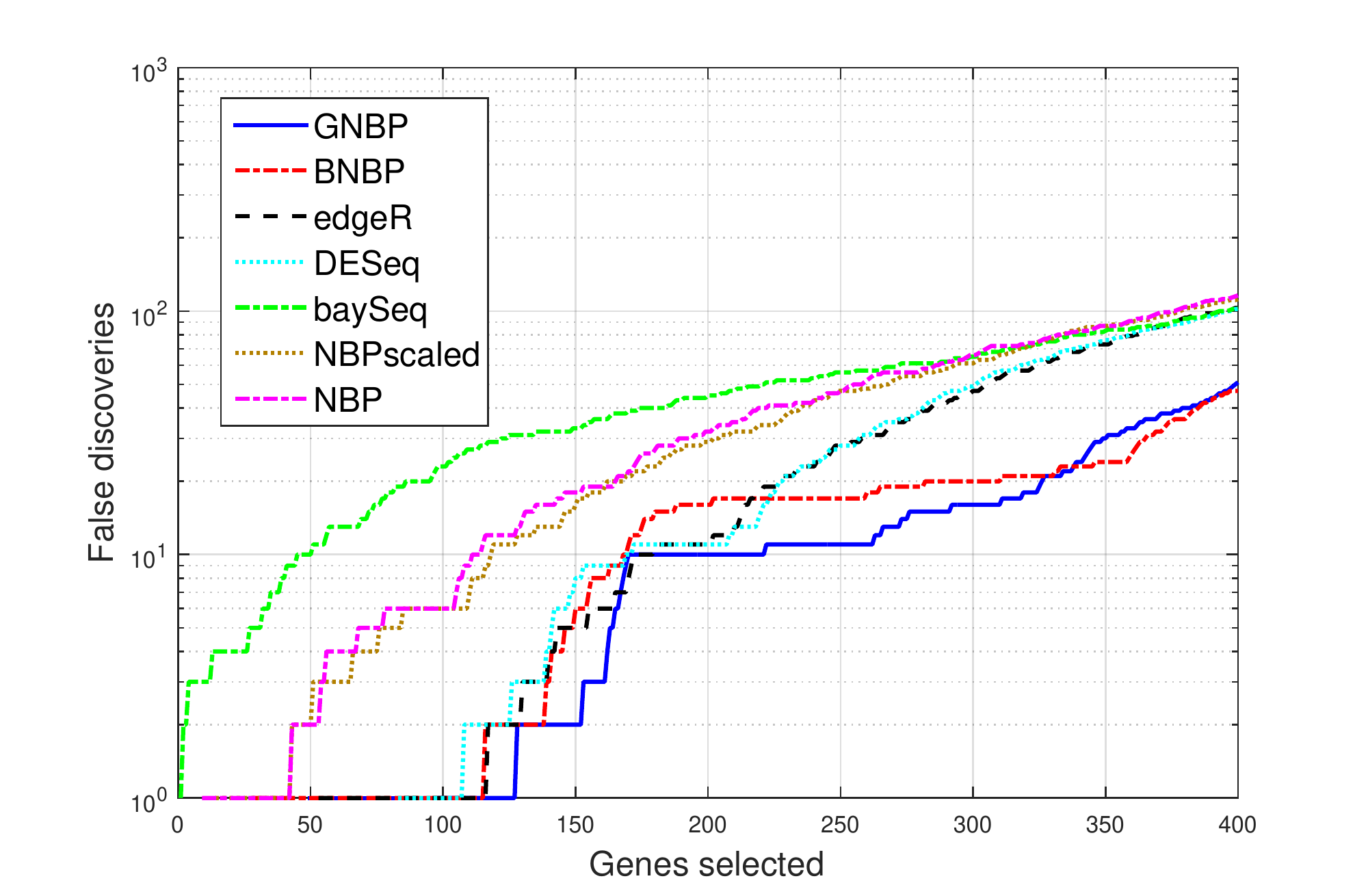}
	\caption{False discovery plots for different methods on the BGI dataset  from the SEQC project, with the log2 cut-off value fixed at 2. The x-axis shows the number of genes selected, in order of their detected differential expression levels, while the y-axis shows the number of selected genes that are false positives.}
	\label{fig:fd}
\end{figure}

\end{document}